\documentclass[12pt,a4paper]{article}
\usepackage[T1]{fontenc}
\usepackage[sc,osf]{mathpazo}
\usepackage{a4wide}  
\usepackage{amsfonts}
\usepackage{amssymb}
\usepackage{amsmath}
\usepackage{bbm}
\usepackage{booktabs} % for much better looking tables
\usepackage{ifpdf}
\ifpdf
\usepackage[pdftex,unicode,implicit]{hyperref}
\hypersetup{%
  pdftitle    = {Non-Abelian, supersymmetric black holes and strings in 5 dimensions}
  pdfkeywords = {gravity, supergravity, black hole, monopole, Wu-Yang, meron, 
multicenter, Yang-Mills,  Einstein-Yang-Mills, Higgs, non-Abelian, singular 
monopole, 't~Hooft-Polyakov monopole, Bogomol'nyi, inspanton,
Belavin-Polyakov-Schwarz-Tyupkin instanton, coloured, Protogenov},
  pdfauthor   = {Patrick Meessen, Tomas Ortin and Pedro Fernandez-Ramirez},
  plainpages  = true,
  colorlinks  = true,
  citecolor   = blue,
  urlcolor    = red,
  linkcolor   = black
}
\newcommand{\hepth}[1]{{\tt
\href{http://www.arXiv.org/abs/hep-th/#1}{hep-th/#1}}}

\newcommand{\arxiv}[1]{{\tt
\href{http://www.arXiv.org/abs/#1}{#1}}}
\else
  \usepackage[dvips]{graphicx}
  \usepackage[unicode,implicit]{hyperref}
  \newcommand{\hepth}[1]{{\tt hep-th/#1}}

  \newcommand{\arxiv}[1]{{\tt arXiv:#1}}
\fi
\usepackage{tikz}\newcommand{\FPAUO}[2]{
\tikz[scale=.13,
         Uniovi/.style={color=green!51!blue, fill=green!51!blue}
 ] {
 \fill[Uniovi] (0,0) circle (10);
 \fill[white] (0,7) circle (1.5);
 \draw[Uniovi] (-2,7.5) rectangle (2,5.5);
 \fill[white] (-0.3,6.6) rectangle (0.3,0);   % 1.7 cm 
 \fill[white] ( -0.9,6.2) rectangle (.9 ,5.6);
 \fill[white] (-1.4, 5.2) rectangle (1.4, 4.6);
 \fill[white] (0,0) ellipse (3.5 and 4);
 \fill[Uniovi] (-2.5,0.3) rectangle (2.5,-0.3);
 \fill[Uniovi] (-2,2.3) rectangle (2,1.7);
 \fill[Uniovi] (-2,-2.3) rectangle (2,-1.7);
 \fill[white] (-4.5,5.5) rectangle (-2.7,4.9);
 \fill[white] (-3.9,6.1) rectangle (-3.3,4.3);
 \fill[white] (4.5,5.5) rectangle (2.7,4.9);
 \fill[white] (3.9,6.1) rectangle (3.3,4.3);
 \foreach \x in { 0,..., 3 }
   \foreach \y in { 0,...,\x}
    {
     \fill[white] (-6-\x*0.7+\y*1.4,3.5-\x *1.97) -- (-5.6-\x*0.7+\y*1.4,2.4-\x *1.97) -- (-6.4-\x*0.7+\y*1.4,2.4-\x *1.97) -- cycle;
     \fill[white] (6-\x*0.7+\y*1.4,3.5-\x *1.97) -- (5.6-\x*0.7+\y*1.4,2.4-\x *1.97) -- (6.4-\x*0.7+\y*1.4,2.4-\x *1.97) -- cycle;
   };
 \draw (0,-6) node[
                               text centered, 
                               color=white, 
                               font={\fontsize{8}{4}\sffamily\selectfont}
                             ] {FPAUO-#1/#2};
}} 
\usepackage{pgfplots}    % This is needed for plotting data
\makeatletter
\@addtoreset{equation}{section}
\makeatother

\begin{document}

\begin{flushright}
\small
\FPAUO{15}{18}\\
IFT-UAM/CSIC-15-088\\
\texttt{arXiv:yymm.nnnn [hep-th]}\\
December 22\textsuperscript{nd}, 2015\\
\normalsize
\end{flushright}

\begin{center}

{\Large {\bf Non-Abelian, supersymmetric black holes and strings in 5 dimensions}}

\vspace{.5cm}

\renewcommand{\thefootnote}{\alph{footnote}}
{\sl\large Patrick Meessen$^{2}$}${}^{,}$\footnote{E-mail: {\tt meessenpatrick [at] uniovi.es}},
{\sl\large Tom\'{a}s Ort\'{\i}n$^{1}$}${}^{,}$\footnote{E-mail: {\tt Tomas.Ortin [at] csic.es}}
{\sl\large and Pedro F.~Ram\'{\i}rez$^{1}$}${}^{,}$\footnote{E-mail: {\tt p.f.ramirez [at]  csic.es}},

\setcounter{footnote}{0}
\renewcommand{\thefootnote}{\arabic{footnote}}

\vspace{1cm}

${}^{1}${\it Instituto de F\'{\i}sica Te\'orica UAM/CSIC\\
  C/ Nicol\'as Cabrera, 13--15, C.U.~Cantoblanco, E-28049 Madrid, Spain}\\
\vspace{0.3cm}

${}^{2}${\it HEP Theory Group, Departamento de F\'{\i}sica, Universidad de Oviedo\\
  Avda.~Calvo Sotelo s/n, E-33007 Oviedo, Spain}\\

\vspace{.5cm}

%%%%%%%%%%%%%%%%%%%%%%%%%%%%%%%%%%%%%%%%%%%%%%%%%%%%%%%%%%%%%%%%%%%%%%

{\bf Abstract}

\end{center}

\begin{quotation}
{\small 

We construct and study the first supersymmetric black-hole and black-string
solutions of non-Abelian-gauged $\mathcal{N}=1,d=5$ supergravity (
$\mathcal{N}=1,d=5$ Super-Einstein-Yang-Mills theory) with non-trivial
SU$(2)$ gauge fields: BPST instantons for black holes and BPS monopoles
of different kinds ('t~Hooft-Polyakov, Wu-Yang and Protogenov) for black
strings and also for certain black holes that are well defined solutions
only for very specific values of all the moduli. Instantons, as well as 
colored monopoles do not contribute to the
masses and tensions but do contribute to the entropies.

The construction is based on the characterization of the supersymmetric solutions of 
gauged $\mathcal{N}=1,d=5$ supergravity coupled to vector multiplets achieved
in Ref.~\cite{Bellorin:2007yp} which we elaborate upon by finding the rules
to construct supersymmetric solutions with one additional isometry, both for
the timelike and null classes. These rules automatically connect the
timelike and null non-Abelian supersymmetric solutions of $\mathcal{N}=1,d=5$
SEYM theory with the timelike ones of $\mathcal{N}=2,d=4$
SEYM theory \cite{Huebscher:2007hj,Hubscher:2008yz} by dimensional reduction
and oxidation. In the timelike-to-timelike case the singular Kronheimer
reduction recently studied in Ref.~\cite{Bueno:2015wva} plays a crucial role.
}
\end{quotation}

\newpage
%%%%%%%%%%%%%%%%%%%%%%%%%%%%%%%%%%%%%%%%%%%%%%%%%%%%%%%%%%%%%%%%%%%%%%
%%%%%%%%%%%%%%%%%%%%%%%%%%%%%%%%%%%%%%%%%%%%%%%%%%%%%%%%%%%%%%%%%%%%%%
%%%%%%%%%%%%%%%%%%%%%%%%%%%%%%%%%%%%%%%%%%%%%%%%%%%%%%%%%%%%%%%%%%%%%%
%%%%%%%%%%%%%%%%%%%%%%%%%%%%%%%%%%%%%%%%%%%%%%%%%%%%%%%%%%%%%%%%%%%%%%
\pagestyle{plain}
%%%%%%%%%%%%%%%%%%%%%%%%%%%%%%%%%%%%%%%%%%%%%%%%%%%%%%%%%%%%%%%%%%%%%%
%%%%%%%%%%%%%%%%%%%%%%%%%%%%%%%%%%%%%%%%%%%%%%%%%%%%%%%%%%%%%%%%%%%%%%
%%%%%%%%%%%%%%%%%%%%%%%%%%%%%%%%%%%%%%%%%%%%%%%%%%%%%%%%%%%%%%%%%%%%%%
%%%%%%%%%%%%%%%%%%%%%%%%%%%%%%%%%%%%%%%%%%%%%%%%%%%%%%%%%%%%%%%%%%%%%%
%%%%%%%%%%%%%%%%%%%%%%%%%%%%%%%%%%%%%%%%%%%%%%%%%%%%%%%%%%%%%%%%%%%%%%

\tableofcontents

%\newpage

%%%%%%%%%%%%%%%%%%%%%%%%%%%%%%%%%%%%%%%%%%%%%%%%%%%%%%%%%%%%%%%%%%%%%%
%%%%%%%%%%%%%%%%%%%%%%%%%%%%%%%%%%%%%%%%%%%%%%%%%%%%%%%%%%%%%%%%%%%%%%
%%%%%%%%%%%%%%%%%%%%%%%%%%%%%%%%%%%%%%%%%%%%%%%%%%%%%%%%%%%%%%%%%%%%%%
%%%%%%%%%%%%%%%%%%%%%%%%%%%%%%%%%%%%%%%%%%%%%%%%%%%%%%%%%%%%%%%%%%%%%%
\section*{Introduction}
%%%%%%%%%%%%%%%%%%%%%%%%%%%%%%%%%%%%%%%%%%%%%%%%%%%%%%%%%%%%%%%%%%%%%%
%%%%%%%%%%%%%%%%%%%%%%%%%%%%%%%%%%%%%%%%%%%%%%%%%%%%%%%%%%%%%%%%%%%%%%
%%%%%%%%%%%%%%%%%%%%%%%%%%%%%%%%%%%%%%%%%%%%%%%%%%%%%%%%%%%%%%%%%%%%%%
%%%%%%%%%%%%%%%%%%%%%%%%%%%%%%%%%%%%%%%%%%%%%%%%%%%%%%%%%%%%%%%%%%%%%%

The search for classical solutions of General Relativity and theories of
gravity in general has proven to be one of the most fruitful approaches to
study this universal and mysterious interaction. This is partially due to the
non-perturbative information they provide, which we do not know how to obtain
otherwise. It is fair to say that some of the solutions discovered (such as
the Schwarzschild and Kerr black-hole solutions, the cosmological ones or the
AdS$_{5}\times$S$^{5}$ solution of type~IIB supergravity) have opened entire
fields of research.

Some of the most interesting solutions are supported by fundamental matter
fields and a large part of the search for gravity solutions has been carried
out in theories in which gravity is coupled to different forms of matter,
usually scalar fields, Abelian vector and $p$-form fields coupled in
gauge-invariant ways among themselves and to scalars, as suggested by
superstring and supergravity theories, for instance. The solutions of gravity
coupled to non-Abelian vector fields have been much less studied because of
the complexity of the equations. Most of the genuinely non-Abelian solutions
found so far, such as the Bartnik-McKinnon particle \cite{Bartnik:1988am} and
its black hole-type generalizations \cite{Volkov:1989fi}, in the SU$(2)$
Einstein-Yang-Mills (EYM) theory, are only known numerically, which makes them
more difficult to study and generalize.

Supersymmetry can simplify dramatically the construction of classical
solutions, providing in some cases recipes to construct systematically whole
families of solutions that have the property of being ``supersymmetric'' or
``having unbroken supersymmetry'', or being ``BPS'' (a much less precise term)
because these solutions satisfy much easier to solve first-order differential
equations.\footnote{For a general review on the construction of supersymmetric
  solutions of supergravity theories, including some of those that we are
  going to study here, see Ref.~\cite{kn:GS}.} These techniques can be applied to
non-supersymmetric theories if we can ``embed'' them in a larger
supersymmetric theory from which they can be obtained by a consistent
truncation that, in particular, gets rid of the fermionic fields.

In order to apply these techniques to the case of theories of gravity coupled
to fundamental matter fields we must embed the theories first in supergravity
theories. $d=4$ EYM theories can be embedded almost trivially in
$\mathcal{N}=1,d=4$ gauged supergravity coupled to vector supermultiplets, but
there are no supersymmetric black-hole or more general particle-like solutions
in $\mathcal{N}=1,d=4$ supergravity: all the supersymmetric solutions of these
theories belong to the null class\footnote{The Killing spinor of the
  supersymmetric solutions in the null (resp.~timelike) class gives rise to a
  null (resp.~timelike) Killing vector bilinear.} and describe, generically,
massless solutions such as gravitational waves and also black strings (whose
tension does not count as a mass). This could well explain why there are no
simple analytic solutions of the EYM theory.

Embedding of $d=4$ EYM theories in extended ($\mathcal{N}>1$) $d=4$
supergravity theories turns out to be impossible, since the latter always
include additional scalar fields charged under the non-Abelian fields which
cannot be consistently truncated away. On the other hand, these scalar fields
(or part of them) can also be interpreted as Higgs fields and we can think of
those supergravities (which we will call Super-Einstein-Yang-Mills (SEYM)
theories) as the minimal supersymmetric generalizations of the
Einstein-Yang-Mills-Higgs (EYMH) theory. Actually, some solutions of the SEYM
theories are also solutions of the EYMH theory, but this is not generically
true and we cannot say that the EYMH theory is embedded in some SEYM theory.

At any rate, analytic supersymmetric solutions of SEYM or more general gauged
supergravity theories should be much easier to find than solutions of the EYM
theory and, at the same time, much more realistic, since we know there are
scalar fields charged under non-Abelian vector fields in Nature.

This expectation turns out to be true. In 1991 Harvey and Liu
\cite{Harvey:1991jr} and in 1997 Chamseddine and Volkov
\cite{Chamseddine:1997nm} found globally regular gravitating monopole
(``global monopole'') solutions to gauged $\mathcal{N}=4$, $d=4$ supergravity,
a theory that can be related to the Heterotic string. In 1994, a 4-dimensional
black-hole solution with non-Abelian hair was obtained by adding stringy
(Heterotic) $\alpha^{\prime}$ corrections to an $a=1$ dilaton black hole
\cite{Kallosh:1994wy}. This solution was singular in the Einstein
frame.\footnote{We will see, though, that it is closely related to the
  4-dimensional black-hole solutions studied in \cite{Bueno:2014mea} and to
  the 5-dimensional ones presented here.} More recently, the timelike
supersymmetric solutions of gauged $\mathcal{N}=2,d=4$ and $\mathcal{N}=1,d=5$
were characterized, respectively, in
Refs.~\cite{Huebscher:2007hj,Meessen:2012sr} and
\cite{Bellorin:2007yp,Bellorin:2008we},\footnote{In the $\mathcal{N}=1,d=5$
  case, the null supersymmetric solutions were characterized as well.} so the
form of all the fields in those solutions is given in terms of a few functions
that satisfy first-order equations. 

In the 4-dimensional case, these first-order equations are straightforward
generalizations of the well-known Bogomol'nyi monopole equations
\cite{Bogomolny:1975de} whose more general static and spherically symmetric
solutions for the gauge group SU$(2)$ were obtained by Protogenov in
Ref.~\cite{Protogenov:1977tq}. Then, the characterization of timelike
supersymmetric solutions was immediately used to construct, apart from global
monopole solutions, the first analytical, regular, static, non-Abelian
black-hole solutions which cannot be considered as pure Abelian embeddings
\cite{Huebscher:2007hj}, showing how the attractor mechanism works in the
non-Abelian setting \cite{Huebscher:2007hj,Hubscher:2008yz}.  Colored black
holes\footnote{Colored black holes have non-Abelian hair but vanishing
  asymptotic charges. The charges must be screened at infinity because they
  contribute to the near-horizon geometry and to the entropy.}  and two-center
non-Abelian solutions were constructed, respectively, in \cite{Meessen:2008kb}
and \cite{Bueno:2014mea} by using, respectively, ``colored monopole'' and
two-center solutions of the Bogomol'nyi equations.

In the $\mathcal{N}=1,d=5$ SEYM case, the characterization obtained in
Refs.~\cite{Bellorin:2007yp,Bellorin:2008we} has not yet been exploited. Doing
so to construct non-Abelian black-hole and black-string solutions is our main
goal in this paper. It is a well-known fact, one that also holds in the
Abelian (ungauged) case that the vector field strengths of the timelike
supersymmetric solutions of these theories are the sum of two pieces, one of
them self-dual in the hyperK\"ahler base space, \textit{i.e.}~an instanton in
the base space. In the non-Abelian case we are interested in, this fact can be
exploited in an obvious way to add non-Abelian hair to black hole solutions.

As we are going to see, it will be convenient to refine the general
characterization obtained in those references to obtain a simpler recipe to
construct supersymmetric solutions with one additional isometry. These
solutions are still general enough and can also be related to the timelike
supersymmetric solutions of $\mathcal{N}=2,d=4$ SEYM. In the
timelike-to-timelike reduction, we recover the relation between self-dual
instantons in hyperK\"ahler spaces with one isometry and BPS monopoles in
$\mathbb{E}^{3}$ found by Kronheimer in Ref.~\cite{kn:KronheimerMScThesis}. As
we have shown in Ref.~\cite{Bueno:2015wva} this redox relation brings us from
singular colored monopoles to globally regular BPST instantons and vice-versa
and it will allow us to obtain regular black holes with a BPST instanton
field. 

The recipes we have obtained can be applied to any model of
$\mathcal{N}=1,d=5$ supergravity coupled to vector multiplets in which a
non-Abelian subgroup of the perturbative duality group can be gauged. The
explicit solutions we will construct will belong to a particular model, the
ST$[2,5]$ model which is the smallest of the ST$[2,n]$ family of models
admitting a SU$(2)$ gauging. These models are consistent truncations of
$\mathcal{N}=1,d=10$ supergravity coupled to a number of vector multiplets on
$T^{5}$ and, for low values of $n$, they can be embedded in Heterotic string
theory. The SU$(2)$ gauging can be associated to the enhancement of symmetry
at the self-dual radius U$(1)\times$U$(1)\rightarrow $U$(1)\times$SU$(2)$,
although, in order to study the details of the embedding of our model in
Heterotic string theory (which will be our next goal) more work will be
necessary.

This paper is organized as follows: in Section~\ref{sec-n1d5} we review the
gauging of a non-Abelian group of isometries of an $\mathcal{N}=1,d=5$
supergravity theory coupled to vector multiplets. The result of this procedure
is what we call an $\mathcal{N}=1,d=5$ Super-Einstein-Yang-Mills (SEYM)
theory. In Section~\ref{sec-susysolutions} we review and extend the results of
Ref.~~\cite{Bellorin:2007yp} on the characterization of the supersymmetric
solutions of $\mathcal{N}=1,d=5$ SEYM theories, giving the recipe to construct
those admitting additional isometries and showing how they are related to the
analogous supersymmetric solutions of $\mathcal{N}=2,d=4$ SEYM theories
characterized in Ref.~\cite{Hubscher:2008yz,Meessen:2012sr}. We will then use
these results in Sec.~\ref{sec-solutions} to construct black holes and black
strings (in the timelike and null cases, respectively) of the SU$(2)$-gauged
ST$[2,5]$ model of $\mathcal{N}=1,d=5$ supergravity and to study their
relations, via dimensional reduction, to the non-Abelian timelike supersymmetric
solutions (black holes and global monopoles) of the SU$(2)$-gauged ST$[2,5]$
model of $\mathcal{N}=2,d=4$ supergravity (see Ref.~\cite{Bueno:2014mea}).
Our conclusions are given in Section~\ref{sec-conclusions}.
Appendix~\ref{ap-reduction} reviews the reduction of ungauged
$\mathcal{N}=1,d=5$ supergravity to a cubic model of $\mathcal{N}=2,d=4$
supergravity, with the relation between the 5- and 4-dimensional fields for
any kind of solution (supersymmetric or not). This relation remains true for
gauged supergravity theories under standard dimensional reduction (which does
not change the gauge group). Finally, Appendix~\ref{sec-Protogenov} review the
spherically-symmetric solutions of the Bogomol'nyi equation in
$\mathbb{E}^{3}$ for SU$(2)$.

%%%%%%%%%%%%%%%%%%%%%%%%%%%%%%%%%%%%%%%%%%%%%%%%%%%%%%%%%%%%%%%%%%%%%%
%%%%%%%%%%%%%%%%%%%%%%%%%%%%%%%%%%%%%%%%%%%%%%%%%%%%%%%%%%%%%%%%%%%%%%
%%%%%%%%%%%%%%%%%%%%%%%%%%%%%%%%%%%%%%%%%%%%%%%%%%%%%%%%%%%%%%%%%%%%%%
%%%%%%%%%%%%%%%%%%%%%%%%%%%%%%%%%%%%%%%%%%%%%%%%%%%%%%%%%%%%%%%%%%%%%%
\section{$\mathcal{N}=1,d=5$ SEYM theories}
\label{sec-n1d5}
%%%%%%%%%%%%%%%%%%%%%%%%%%%%%%%%%%%%%%%%%%%%%%%%%%%%%%%%%%%%%%%%%%%%%%
%%%%%%%%%%%%%%%%%%%%%%%%%%%%%%%%%%%%%%%%%%%%%%%%%%%%%%%%%%%%%%%%%%%%%%
%%%%%%%%%%%%%%%%%%%%%%%%%%%%%%%%%%%%%%%%%%%%%%%%%%%%%%%%%%%%%%%%%%%%%%
%%%%%%%%%%%%%%%%%%%%%%%%%%%%%%%%%%%%%%%%%%%%%%%%%%%%%%%%%%%%%%%%%%%%%%

In this section we give a brief description of general $\mathcal{N}=1,d=5$
Super-Einstein-Yang-Mills (SEYM) theories. These are theories of
$\mathcal{N}=1,d=5$ supergravity coupled to $n_{v}$ vector supermultiplets (no
hypermultiplets) in which a necessarily non-Abelian group of isometries of the
Real Special manifold has been gauged. These theories can be considered the
simplest supersymmetrization of non-Abelian Einstein-Yang-Mills theories in
$d=5$.  Our conventions are those in
Refs.~\cite{Bellorin:2006yr,Bellorin:2007yp} which are those of
Ref.~\cite{Bergshoeff:2004kh} with minor modifications.

The supergravity multiplet is constituted by the graviton $e^{a}{}_{\mu},$ the
gravitino $\psi_{\mu}^{i}$ and the graviphoton $A_{\mu}$.  All the spinors are
symplectic Majorana spinors and carry a fundamental $SU(2)$ R-symmetry index.
The $n_{v}$ vector multiplets, labeled by $x=1,....,n_{v}$ consist of a real
vector field $A^{x}{}_{\mu},$ a real scalar $\phi^{x}$ and a gaugino
$\lambda^{i\, x}$. 

The full theory is formally invariant under a $SO(n_{v}+1)$ group\footnote{The
  theory will only be invariant under a subgroup of $SO(n_{v}+1)$.} that mixes
the matter vector fields $A^{x}{}_{\mu}$ with the graviphoton $A_{\mu}\equiv
A^{0}{}_{\mu}$ and it is convenient to combine them into an $SO(n_{v}+1)$
vector $(A^{I}{}_{\mu})= (A^{0}{}_{\mu},A^{x}{}_{\mu})$. It is also convenient
to define a $SO(n_{v}+1)$ vector of functions of the scalars
$h^{I}(\phi)$. These $n_{v}+1$ functions of $n_{v}$ scalar must satisfy a
constraint. $\mathcal{N}=1,d=5$ supersymmetry determines that this constraint is of the
form

\begin{equation}
\label{eq:Ch3=1}
C_{IJK}h^{I}(\phi)h^{J}(\phi)h^{K}(\phi)=1,
\end{equation}

\noindent
where the constant symmetric tensor $C_{IJK}$ completely characterizes the
theory and the \textit{Special Real geometry} of the scalar manifold. In
particular, the kinetic matrix of the vector fields $a_{IJ}(\phi)$ and the
metric of the scalar manifold $g_{xy}(\phi)$ can be derived from it as
follows: first, we define
\begin{equation}
\label{eq:h_I}
h_{I}\equiv C_{IJK}h^{J}h^{K},
\,\,\,\,\,
\Rightarrow
\,\,\,\,\,
h^{I}h_{I}=1,
\end{equation}

\noindent
and 

\begin{equation}
h^{I}_{x} 
\equiv
-\sqrt{3} h^{I}{}_{,x}
\equiv  
-\sqrt{3} \frac{\partial h^{I}}{\partial\phi^{x}},  
\hspace{1cm}
h_{Ix}
\equiv  
+\sqrt{3}h_{I, x},
\,\,\,\,\,
\Rightarrow
\,\,\,\,\,
h_{I}h^{I}_{x}
=
h^{I}h_{Ix}
=
0.   
\end{equation}

\noindent
Then, $a_{IJ}$ is defined implicitly by the relations

\begin{equation}
h_{I}  = a_{IJ}h^{I},
\hspace{1cm}
h_{Ix}  = a_{IJ}h^{J}{}_{x}.
\end{equation}

\noindent
It can be checked that

\begin{equation}
a_{IJ}
=
-2C_{IJK}h^{K} +3h_{I}h_{J}.  
\end{equation}

The metric of the scalar manifold $g_{xy}(\phi)$, which we will use to raise
and lower $x,y$ indices is (proportional to) the pullback of $a_{IJ}$

\begin{equation}
g_{xy}
\equiv
a_{IJ}h^{I}{}_{x}h^{J}{}_{y}
=
-2C_{IJK}h_{x}^{I}h_{y}^{J}h^{K}.
\end{equation}

The functions $h^{I}$ and their derivatives $h^{I}_{x}$ satisfy the following completeness relation:

\begin{equation}
\label{eq:completeness}
a_{IJ} 
=
h_{I}h_{J}+g_{xy}h^{x}_{I}h^{y}_{J}.  
\end{equation}

By assumption, the real Real Special structure is invariant under
reparametrizations generated by vectors $k_{I}{}^{x}(\phi)$\footnote{Some of
these vectors may be identically zero. This is price paid for labeling the
gauge vectors and the Killing vectors with the same indices.}

\begin{equation}
\delta\phi^{x} = c^{I}k_{I}{}^{x},
\end{equation}

\noindent
satisfying the Lie algebra\footnote{Some of the structure constants may vanish
  identically, but it is assumed that some of them do not because, otherwise,
  we would be dealing with an ungauged supergravity.}

\begin{equation}
[k_{I},k_{J}]= -f_{IJ}{}^{K}k_{K}.
\end{equation}

\noindent
The invariance of the metric $g_{xy}$ implies that the vectors
$k_{I}{}^{x}(\phi)$ are Killing vectors. The invariance of the constraint
Eq.~(\ref{eq:Ch3=1}) implies the invariance of the $C_{IJK}$
tensor

\begin{equation}
\label{eq:CIJKinvariance}
-3f_{I(J}{}^{M}C_{KL)M}=0.   
\end{equation}

\noindent
Multiplying this identity by $h^{J}h^{K}h^{L}$ we get another
important relation:

\begin{equation}
\label{eq:fh2=0}
f_{IJ}{}^{K}h^{J}h_{K}=0.  
\end{equation}

The functions $h^{I}(\phi)$, in their turn, must be invariant up to
$SO(n_{v}+1)$ rotations, that is

\begin{equation}
\label{eq:kfh}
k_{I}{}^{x}\partial_{x}h^{J} -f_{IK}{}^{J}h^{K}=0,
\,\,\,\,\,
\Rightarrow
\,\,\,\,\,
k_{I}{}^{x}= -\sqrt{3}f_{IJ}{}^{K}h_{K}^{x}h^{J},
\,\,\,\,\,
\Rightarrow
\,\,\,\,\,
h^{I}k_{I}{}^{x}
=
0,
\end{equation}

\noindent
where we have used the completeness relation
Eq.~(\ref{eq:completeness}) and Eq.~(\ref{eq:fh2=0}).

If the real special manifold is a symmetric space, then the tensor $C_{IJK}$
satisfies the identity

\begin{equation}
C^{IJK}C_{J(LM}C_{NP)K}=\tfrac{1}{27}\delta^{I}{}_{(L}C_{MNP)}\, ,  
\end{equation}

\noindent
where $C^{IJK}=C_{IJK}$. In these spaces we can solve immediately $h^{I}$ in
terms of the $h_{I}$

\begin{equation}
\label{eq:stabilizationsymmetric}
h^{I}=27 C^{IJK}h_{J}h_{K}\, ,
\,\,\,\,\,
\Rightarrow
\,\,\,\,\,
C^{IJK}h_{I}h_{J}h_{K} = \frac{1}{27}\, .  
\end{equation}

To gauge this global symmetry group we promote the constant parameters $c^{I}$
to arbitrary spacetime functions identifying them with the gauge parameters of
the vector fields $\Lambda^{I}(x)$ $c^{I}\rightarrow -g\Lambda^{I}(x)$. The
gauge transformations scalars $\phi^{x}$, the functions $h^{I}$ and the
$A^{I}{}_{\mu}$ take the form

\begin{eqnarray}
\label{eq:deltaphix}
\delta_{\Lambda}\phi^{x} 
& = & 
-g\Lambda^{I} k_{I}{}^{x},
\\
& & \nonumber \\
\label{eq:deltahi}
\delta_{\Lambda} h^{I} 
& = &
-gf_{JK}{}^{I}\Lambda^{J} h^{K},
\\
& & \nonumber \\
\delta_{\Lambda}A^{I}{}_{\mu}
& = & 
\partial_{\mu}\Lambda^{I}
+gf_{JK}{}^{I}A^{J}{}_{\mu}\Lambda^{K}
\equiv
\mathfrak{D}_{\mu}\Lambda^{I},
\end{eqnarray}

\noindent
where $\mathfrak{D}_{\mu}$ is the gauge-covariant
derivative. $\mathfrak{D}_{\mu}h^{I}$ has the same expression as
$\mathfrak{D}_{\mu}\Lambda^{I}$ and have the same gauge transformations as
$h^{I}$ and $\Lambda^{I}$.  We also have
\begin{eqnarray}
\mathfrak{D}_{\mu} h_{I}
& = & 
\partial_{\mu} h_{I} +gf_{IJ}{}^{K} A^{J}{}_{\mu} h_{K},
\\
& & \nonumber \\
\mathfrak{D}_{\mu}C_{IJK}
& = &
0.  
\end{eqnarray}

On the other hand, the gauge-covariant derivative of the scalars is given
by

\begin{equation}
\mathfrak{D}_{\mu} \phi^{x} 
= 
\partial_{\mu} \phi^{x}+gA^{I}{}_{\mu} k_{I}{}^{x},
\end{equation}

\noindent
and transforms as

\begin{equation}
\delta_{\Lambda} \mathfrak{D}_{\mu} \phi^{x}
=
-g\Lambda^{I}\partial_{y}k_{I}{}^{x}\mathfrak{D}_{\mu}\phi^{x}. 
\end{equation}

The gauginos $\lambda^{i\, x}$ transform in exactly the same way as
$\mathfrak{D} \phi^{x}$ and their gauge-covariant derivatives are
identical to the second covariant derivative of $\phi^{x}$:

\begin{equation}
\mathfrak{D}_{\mu} \mathfrak{D}_{\nu}\phi^{x} 
 = 
\partial_{\mu}\mathfrak{D}_{\nu}\phi^{x}
-\Gamma_{\mu\nu}^{\rho}\mathfrak{D}_{\rho}\phi^{x} 
+\Gamma_{yz}{}^{x}\mathfrak{D}_{\mu}\phi^{y}\mathfrak{D}_{\nu}\phi^{z}
+ gA^{I}{}_{\mu}\partial_{y}k_{I}{}^{x}
\mathfrak{D}_{\nu}\phi^{y}.
\end{equation}

The gauge-covariant vector field strength has the standard form

\begin{equation}
F^{I}{}_{\mu\nu}=2\partial_{[\mu}A^{I}{}_{\nu]}  
+gf_{JK}{}^{I}A^{J}{}_{\mu}A^{K}{}_{\nu}.
\end{equation}

The bosonic action of $\mathcal{N}=1,d=5$ SEYM is given in terms of
$a_{IJ},g_{xy},C_{IJK}$ and the structure constants $f_{IJ}{}^{K}$ by

\begin{equation}
\begin{array}{rcl}
S & = &  {\displaystyle\int} d^{5}x\sqrt{g}\
\biggl\{
R
+{\textstyle\frac{1}{2}}g_{xy}\mathfrak{D}_{\mu}\phi^{x}
\mathfrak{D}^{\mu}\phi^{y}
-{\textstyle\frac{1}{4}} a_{IJ} F^{I\, \mu\nu}F^{J}{}_{\mu\nu}
+\tfrac{1}{12\sqrt{3}}C_{IJK}
{\displaystyle\frac{\varepsilon^{\mu\nu\rho\sigma\alpha}}{\sqrt{g}}}
\left[
F^{I}{}_{\mu\nu}F^{J}{}_{\rho\sigma}A^{K}{}_{\alpha}
\right.
\\ \\ & & 
\left.
-\tfrac{1}{2}gf_{LM}{}^{I} F^{J}{}_{\mu\nu} 
A^{K}{}_{\rho} A^{L}{}_{\sigma} A^{M}{}_{\alpha}
+\tfrac{1}{10} g^{2} f_{LM}{}^{I} f_{NP}{}^{J} 
A^{K}{}_{\mu} A^{L}{}_{\nu} A^{M}{}_{\rho} A^{N}{}_{\sigma} A^{P}{}_{\alpha}
\right]
\biggr\}.
\end{array}
\end{equation}

\noindent
Observe that this action does not contain a scalar potential
$V(\phi)$ because 

\begin{equation}
V(\phi) 
=
\tfrac{3}{2}g^{2}h^{I}h^{J} 
k_{I}{}^{x} k_{J}{}^{y} g_{xy}\, ,
\end{equation}

\noindent
(the expression that follows from the general formula in
Ref.~\cite{Bergshoeff:2004kh}) vanishes identically for the kind of gaugings
considered here, owing to the property Eq.~(\ref{eq:kfh}). This fact is
associated to the vanishing of the corresponding fermion shift in the
gauginos' supersymmetry transformations.

The equations of motion for the bosonic fields are

\begin{eqnarray}
\mathcal{E}_{\mu\nu} 
& \equiv &
\frac{1}{2\sqrt{g}} e_{a(\mu}
\frac{\delta S}{\delta e_{a}{}^{\nu)}}
\nonumber \\
& & \nonumber \\
& = &
G_{\mu\nu}
-{\textstyle\frac{1}{2}}a_{IJ}\left(F^{I}{}_{\mu}{}^{\rho} F^{J}{}_{\nu\rho}
-{\textstyle\frac{1}{4}}g_{\mu\nu}F^{I\, \rho\sigma}F^{J}{}_{\rho\sigma}
\right)      
\nonumber\\
& & \nonumber  \\
& & 
+{\textstyle\frac{1}{2}}g_{xy}\left(\mathfrak{D}_{\mu}\phi^{x} 
\mathfrak{D}_{\nu}\phi^{y}
-{\textstyle\frac{1}{2}}g_{\mu\nu}
\mathfrak{D}_\rho\phi^{x} \mathfrak{D}^{\rho}\phi^{y}\right)
\\ 
& & \nonumber \\
\mathcal{E}_{I}{}^{\mu} 
& \equiv &
\frac{1}{\sqrt{g}}\frac{\delta S}{\delta A^{I}{}_{\mu}}
\nonumber \\
& & \nonumber \\
& = & 
\mathfrak{D}_{\nu}\left(a_{IJ} F^{J\, \nu\mu}\right)
+{\textstyle\frac{1}{4\sqrt{3}}} 
\frac{\varepsilon^{\mu\nu\rho\sigma\alpha}}{\sqrt{g}}
C_{IJK} F^{J}{}_{\nu\rho}F^{k}{}_{\sigma\alpha}
+g k_{I\,x} \mathfrak{D}^{\mu}\phi^{x}
\\
& & \nonumber \\
\mathcal{E}^{x} 
& \equiv &
-\frac{g^{xy}}{\sqrt{g}}
\frac{\delta S}{\delta \phi^{y}}
\nonumber \\
& & \nonumber \\
& = & 
\mathfrak{D}_{\mu}\mathfrak{D}^{\mu}\phi^{x} 
+{\textstyle\frac{1}{4}}g^{xy} \partial_{y}
a_{IJ} F^{I\, \rho\sigma}F^{J}{}_{\rho\sigma}.
\end{eqnarray}

The supersymmetry transformation rules for the bosonic fields are

\begin{equation}
\label{eq:N1d5ungaugedsugrabosonsusyrules}
\begin{array}{rcl}
\delta_{\epsilon} e^{a}{}_{\mu} 
& = & 
{\textstyle\frac{i}{2}} \bar{\epsilon}_{i}\gamma^{a}\psi^{i}_{\mu},
\\
& & \\ 
\delta_{\epsilon} A^{I}{}_{\mu} 
& = & 
-{\textstyle\frac{i\sqrt{3}}{2}}h^{I}\bar{\epsilon}_{i}\psi^{i}_{\mu}
+{\textstyle\frac{i}{2}}h^{I}_{x}\bar{\epsilon}_{i}\gamma_{\mu}\lambda^{i\, x}
,
\\
& & \\
\delta_{\epsilon} \phi^{x} 
& = & 
{\textstyle\frac{i}{2}}\bar{\epsilon}_{i}\lambda^{i\, x}.
\end{array}
\end{equation}

\noindent
and the corresponding transformation rules for the fermionic fields evaluated
on vanishing fermions are

\begin{eqnarray}
  \delta_{\epsilon}\psi^{i}_{\mu} 
  & = & 
  \nabla_{\mu}\epsilon^{i}
  -{\textstyle\frac{1}{8\sqrt{3}}}h_{I}F^{I\,\alpha\beta}
  \left(\gamma_{\mu\alpha\beta}-4g_{\mu\alpha}\gamma_{\beta}\right)
  \epsilon^{i},
  \\
  & & \nonumber \\ 
  \delta_{\epsilon}\lambda^{i\, x} 
  & = &  
  {\textstyle\frac{1}{2}}\left(\not\!\!\mathfrak{D}\phi^{x} 
    -{\textstyle\frac{1}{2}}h^{x}_{I}\not\!F^{I}\right)\epsilon^{i},
\end{eqnarray}

\noindent
where $\nabla_{\mu}\epsilon^{i}$ is just the Lorentz-covariant
derivative on the spinors, given in our conventions by

\begin{equation}
\nabla_{\mu}\epsilon^{i}
=
\left(\partial_{\mu}-\tfrac{1}{4}\not\!\omega_{\mu}\right)
\epsilon^{i}.
\end{equation}

The equations of motion and the supersymmetry transformation rules are the
straightforward covariantization of those of the ungauged theory, except for
the addition of a source to the Maxwell equations corresponding to the charge
carried by the scalar fields.

%%%%%%%%%%%%%%%%%%%%%%%%%%%%%%%%%%%%%%%%%%%%%%%%%%%%%%%%%%%%%%%%%%%%%%
%%%%%%%%%%%%%%%%%%%%%%%%%%%%%%%%%%%%%%%%%%%%%%%%%%%%%%%%%%%%%%%%%%%%%%
%%%%%%%%%%%%%%%%%%%%%%%%%%%%%%%%%%%%%%%%%%%%%%%%%%%%%%%%%%%%%%%%%%%%%%
%%%%%%%%%%%%%%%%%%%%%%%%%%%%%%%%%%%%%%%%%%%%%%%%%%%%%%%%%%%%%%%%%%%%%%
\section{The supersymmetric solutions of $\mathcal{N}=1,d=5$ SEYM theories}
\label{sec-susysolutions}
%%%%%%%%%%%%%%%%%%%%%%%%%%%%%%%%%%%%%%%%%%%%%%%%%%%%%%%%%%%%%%%%%%%%%%
%%%%%%%%%%%%%%%%%%%%%%%%%%%%%%%%%%%%%%%%%%%%%%%%%%%%%%%%%%%%%%%%%%%%%%
%%%%%%%%%%%%%%%%%%%%%%%%%%%%%%%%%%%%%%%%%%%%%%%%%%%%%%%%%%%%%%%%%%%%%%
%%%%%%%%%%%%%%%%%%%%%%%%%%%%%%%%%%%%%%%%%%%%%%%%%%%%%%%%%%%%%%%%%%%%%%

In this section we are going to review first the results of
Ref.~\cite{Bellorin:2007yp} particularized to the case in which there are no
hypermultiplets nor Fayet-Iliopoulos terms. We will simply focus on the final
characterization of the supersymmetric solutions. Then, we will analyze the
form of the solutions that admit an additional isometry and can, therefore, be
dimensionally reduced to $d=4$, following
Refs.~\cite{Gauntlett:2002nw,Bellorin:2006yr}.

Let us start by reminding the reader that a solution of one of the $\mathcal{N}=1,d=5$
SEYM theories is said supersymmetric if the so-called \textit{Killing spinor
  equations}

\begin{equation}
\delta_{\epsilon}\psi^{i}_{\mu} = 0\, ,
\hspace{1cm}
\delta_{\epsilon}\lambda^{i\, x}  = 0\, ,  
\end{equation}

\noindent
written in the background of the solution can be solved for at least one
spinor $\epsilon^{i}(x)$, which is then called \textit{Killing spinor}.  The
supersymmetric solutions of these theories can be classified according to the
causal nature of the Killing vector that one can construct as a bilinear of
the Killing spinor $V^{a}=i\bar{\epsilon}_{i}\gamma^{a}\epsilon^{i}$ as
\textit{timelike} ($V^{a}V_{a}> 0$) or \textit{null} ($V^{a}V_{a}= 0$). These
two cases must be discussed separately.

%%%%%%%%%%%%%%%%%%%%%%%%%%%%%%%%%%%%%%%%%%%%%%%%%%%%%%%%%%%%%%%%%%%%%%
%%%%%%%%%%%%%%%%%%%%%%%%%%%%%%%%%%%%%%%%%%%%%%%%%%%%%%%%%%%%%%%%%%%%%%
%%%%%%%%%%%%%%%%%%%%%%%%%%%%%%%%%%%%%%%%%%%%%%%%%%%%%%%%%%%%%%%%%%%%%%
%%%%%%%%%%%%%%%%%%%%%%%%%%%%%%%%%%%%%%%%%%%%%%%%%%%%%%%%%%%%%%%%%%%%%%
\subsection{Timelike supersymmetric solutions}
%%%%%%%%%%%%%%%%%%%%%%%%%%%%%%%%%%%%%%%%%%%%%%%%%%%%%%%%%%%%%%%%%%%%%%
%%%%%%%%%%%%%%%%%%%%%%%%%%%%%%%%%%%%%%%%%%%%%%%%%%%%%%%%%%%%%%%%%%%%%%
%%%%%%%%%%%%%%%%%%%%%%%%%%%%%%%%%%%%%%%%%%%%%%%%%%%%%%%%%%%%%%%%%%%%%%
%%%%%%%%%%%%%%%%%%%%%%%%%%%%%%%%%%%%%%%%%%%%%%%%%%%%%%%%%%%%%%%%%%%%%%

The fields of the timelike supersymmetric solutions of $\mathcal{N}=1,d=5$ SEYM theories
are completely determined by 

\begin{enumerate}
\item A choice of 4-dimensional (obviously Euclidean) hyperK\"ahler metric

  \begin{equation}
    d\hat{s}^{2}=h_{\underline{m}\underline{n}}(x)dx^{m} dx^{n}\, .
\end{equation}

\noindent
Fields and operators defined in this space are customarily hatted.

\item Vector fields defined in the hyperK\"ahler space, $\hat{A}^{I}$, such
  that their 2-form field strengths, $\hat{F}^{I}(\hat{A})$ are self-dual

\begin{equation}
\label{eq:instantonequation}
\hat{\star}\hat{F}^{I}=+\hat{F}^{I}\, ,  
\end{equation}
  
\noindent
with respect to the hyperK\"ahler metric. This implies that $\hat{A}^{I}$
defines an instanton solution of the Yang-Mills equations in the hyperK\"ahler
space.

\item A set of functions in the hyperK\"ahler space $\hat{f}_{I}$ satisfying
  the equation\footnote{The coefficient of the second term is wrong by a
    factor of 2 in Refs.~\cite{Bellorin:2006yr,Bellorin:2007yp} although all
    subsequent formulae are correct.}

\begin{equation}
\label{eq:fIequation}
  \hat{\mathfrak{D}}{}^{2}\hat{f}_{I} 
  -\tfrac{1}{6}C_{IJK}\hat{F}^{J}\cdot \hat{F}^{K} 
  =
  0\, .
\end{equation}

\end{enumerate}

Given $h_{\underline{m}\underline{n}},\hat{A}^{I}, \hat{f}_{I}$, the physical
fields can be reconstructed as follows:

\begin{enumerate}
\item The functions $\hat{f}_{I}$ are proportional to the $h_{I}(\phi)$
  defined in Eq.~(\ref{eq:h_I}). The proportionality coefficient is called
  $1/\hat{f}$:

  \begin{equation}
\label{eq:h_I/f}
h_{I}/\hat{f} =  \hat{f}_{I}\, .
  \end{equation}

  The functions $h_{I}(\phi)$ satisfy a model-dependent constraint (analogous
  to the constraint satisfied by the functions $h^{I}(\phi)$,
  Eq.~(\ref{eq:Ch3=1})). This constraint can be obtained by solving
  Eq.~(\ref{eq:h_I}) for the $h^{I}$ and substituting the result into
  Eq.~(\ref{eq:Ch3=1}). Therefore, the constraint has the form $F(h_{\cdot})=1$
  where $F$ is a function homogeneous of degree $3/2$ in the $h_{I}$ and,
  substituting the above equation, one gets

  \begin{equation}
  \hat{f}^{-3/2} = F(\hat{f}_{\cdot})\, .  
  \end{equation}

Using this result in Eq.~(\ref{eq:h_I/f}) one gets all the $h_{I}$ as in
terms of the $\hat{f}_{I}$
 
\begin{equation}
 h_{I} = \hat{f}_{I} F^{-2/3}(\hat{f}_{\cdot})\, ,
\end{equation}

\noindent
and, using the expression of the $h^{I}$ in terms of the $h_{I}$, one also
gets the $h^{I}$ in terms of the functions $\hat{f}_{I}$.

If the real special scalar manifold is symmetric, then we can use
Eq.~(\ref{eq:stabilizationsymmetric}) to get

\begin{equation}
\label{eq:f3symmetric}
\hat{f}^{-3}
=
27C^{IJK}\hat{f}_{I}\hat{f}_{J}\hat{f}_{K}\, .  
\end{equation}

\item The scalar fields $\phi^{x}$ can be obtained by inverting the functions
  $h_{I}(\phi)$ or $h^{I}(\phi)$. A parametrization which is always available
  is

\begin{equation}
  \phi^{x}= h_{x}/h_{0}= \hat{f}_{x}/\hat{f}_{0}\, .    
\end{equation}

\item Next, we define the 1-form $\hat{\omega}$ through the equation

\begin{equation}
\label{eq:omegaequation}
\left(\hat{f} d\hat{\omega} \right)^{+}
=
\frac{\sqrt{3}}{2}h_{I}\hat{F}^{I+}\, .
\end{equation}

\item Having solved the above equation for $\hat{\omega}$ we have determined
  completely the metric of the timelike supersymmetric solutions, which is
  given by

\begin{equation}
ds^{2} = \hat{f}^{\, 2}(dt+\hat{\omega})^{2}
-\hat{f}^{\, -1}h_{\underline{m}\underline{n}}dx^{m} dx^{n}\, ,  
\end{equation}

\item Also, the complete 5-dimensional vector fields are given by

\begin{equation}
\label{eq:completevectorfields}
A^{I} = -\sqrt{3}h^{I}e^{0} +\hat{A}^{I}\, ,    
\,\,\,\,\,
\mbox{where}
\,\,\,\,\,
e^{0} 
\equiv
\hat{f} (dt +\hat{\omega})\, ,
\end{equation}

\noindent
so that the spatial components are 

\begin{equation}
A^{I}{}_{\underline{m}} 
= 
\hat{A}^{I}{}_{\underline{m}} -\sqrt{3}h^{I}\hat{f} \hat{\omega}_{\underline{m}}\, .  
\end{equation}

The field strength can be written in the form

\begin{equation}
F^{I}
= 
-\sqrt{3} \hat{\mathfrak{D}}( h^{I} e^{0})  +\hat{F}^{I}\, ,
\end{equation}

\noindent
where $\hat{\mathfrak{D}}$ is the covariant derivative in the hyperK\"ahler
space with connection $\hat{A}^{I}$.

\end{enumerate}

%%%%%%%%%%%%%%%%%%%%%%%%%%%%%%%%%%%%%%%%%%%%%%%%%%%%%%%%%%%%%%%%%%%%%%
%%%%%%%%%%%%%%%%%%%%%%%%%%%%%%%%%%%%%%%%%%%%%%%%%%%%%%%%%%%%%%%%%%%%%%
%%%%%%%%%%%%%%%%%%%%%%%%%%%%%%%%%%%%%%%%%%%%%%%%%%%%%%%%%%%%%%%%%%%%%%
%%%%%%%%%%%%%%%%%%%%%%%%%%%%%%%%%%%%%%%%%%%%%%%%%%%%%%%%%%%%%%%%%%%%%%
\subsubsection{Timelike supersymmetric solutions with one isometry}
%%%%%%%%%%%%%%%%%%%%%%%%%%%%%%%%%%%%%%%%%%%%%%%%%%%%%%%%%%%%%%%%%%%%%%
%%%%%%%%%%%%%%%%%%%%%%%%%%%%%%%%%%%%%%%%%%%%%%%%%%%%%%%%%%%%%%%%%%%%%%
%%%%%%%%%%%%%%%%%%%%%%%%%%%%%%%%%%%%%%%%%%%%%%%%%%%%%%%%%%%%%%%%%%%%%%
%%%%%%%%%%%%%%%%%%%%%%%%%%%%%%%%%%%%%%%%%%%%%%%%%%%%%%%%%%%%%%%%%%%%%%

We are particularly interested in the supersymmetric solutions that have an
additional isometry. Following Refs.~\cite{Gauntlett:2002nw,Gauntlett:2004qy}
we assume that the additional isometry is a triholomorphic isometry of the
hyperK\"ahler metric (\textit{i.e.}~an isometry respecting the hyperK\"ahler
structure), in which case, as shown in Ref.~\cite{Gibbons:1987sp} it must be a
Gibbons-Hawking multi-instanton metric \cite{Gibbons:1979zt}. Assuming $z$ is
the coordinate associated to the additional isometry, these metrics can always
be written in the form

\begin{equation}
\label{eq:GGH}
h_{\underline{m}\underline{n}}dx^{m}dx^{n} 
=  
H^{-1} (dz +\chi)^{2}
+H dx^{r}dx^{r}\, ,\,\,\,\, r=1,2,3\, ,
\end{equation}

\noindent
where the $z$-independent function $H$  and  1-form
$\chi=\chi_{\underline{r}}dx^{r}$ are related by

\begin{equation}
d\chi = \star_{3}dH\, ,  
\end{equation}

\noindent
$\star_{3}$ being the Hodge operator in $\mathbb{E}^{3}$. Assuming now that
the rest of the bosonic fields of the timelike supersymmetric solutions are
$z$-independent one can simplify
Eqs.~(\ref{eq:instantonequation}),(\ref{eq:fIequation}) and
(\ref{eq:omegaequation}).

Let us start with Eq.~(\ref{eq:instantonequation}) and let us assume that the
selfduality of $\hat{F}^{I}$ has been defined with respect to the frame and
orientation

\begin{equation}
\label{eq:Kframe}
\hat{e}^{\, z} = H^{-1/2}(dz+\chi)\, ,
\hspace{1cm}  
\hat{e}^{\, r} = H^{1/2}\delta^{r}{}_{\underline{r}} dx^{r}\, ,
\hspace{1cm}  
\varepsilon_{z123}=+1\, .
\end{equation}

\noindent
Then, following Kronheimer \cite{kn:KronheimerMScThesis},\footnote{See also
  Ref.~\cite{Bueno:2015wva}.}  Eq.~(\ref{eq:instantonequation}) can be
rewritten as Bogomol'nyi equations for a Yang-Mills-Higgs (YMH) system in the
BPS limit in $\mathbb{E}^{3}$ \cite{Bogomolny:1975de}

\begin{equation}
\label{eq:B}
\breve{\mathfrak{D}}_{r}\Phi^{I} = \tfrac{1}{2}\varepsilon_{rst}\breve{F}^{I}{}_{st}\, ,
\end{equation}

\noindent
where the 3-dimensional Higgs field and the vector fields are given
by\footnote{We have rescaled the 3-dimensional fields by a factor of
  $-1/(2\sqrt{6})$ to conform to the normalization of the fields in
  $\mathcal{N}=2,d=4$ supergravity. See Appendix~\ref{ap-reduction}.}

\begin{equation}
\label{eq:Kformulae}
\begin{array}{rcl}
2\sqrt{6}\Phi^{I}
& \equiv &
H\hat{A}^{I}{}_{\underline{z}}\, ,
\\
& & \\
2\sqrt{6}\breve{A}_{\underline{r}}
& \equiv &
-\hat{A}^{I}{}_{\underline{r}}
+\chi_{\underline{r}}\hat{A}^{I}{}_{\underline{z}}\, .
\end{array}
\end{equation}

Thus, we can always construct a selfdual YM instanton in a Gibbons-Hawking
space from a (monopole) solution of the Bogomol'nyi equation of a YMH system
in $\mathbb{E}^{3}$ $(\Phi^{I},\breve{A}^{I}{}_{\underline{r}})$
\cite{kn:KronheimerMScThesis}. Many solutions of these equations are known,
specially in the spherically symmetric case\footnote{see
  Ref.~\cite{Protogenov:1977tq} for the SU$(2)$ case and
  Ref.~\cite{Meessen:2015nla} and references therein for more general gauge
  groups.}. In Ref.~\cite{Bueno:2015wva} this relation has been explored
precisely for the SU$(2)$ monopoles and instantons we are interested in, and
we will make use of those results later.

We can now use this result into Eq.~(\ref{eq:fIequation}), rewriting the
4-dimensional gauge vector in terms of the 3-dimensional gauge vector and
Higgs field defined above and using the harmonicity of $H$ and the Bogomol'nyi
equation to get rid of $\breve{F}^{I}$ and $\breve{\mathfrak{D}}^{2}\Phi^{I}$
(which vanishes identically). The result is the equation in $\mathbb{E}^{3}$

\begin{equation}
\breve{\mathfrak{D}}^{2}\hat{f}_{I} 
-g^{2} f_{IJ}{}^{L}f_{KL}{}^{M}\Phi^{J}\Phi^{K}\hat{f}_{M} 
-8 C_{IJK}\breve{\mathfrak{D}}^{2}\left(\Phi^{J}\Phi^{K}/H\right)  
=
0\, .
\end{equation}

Defining 

\begin{equation}
\hat{f}_{I}
\equiv 
L_{I}+8C_{IJK}\Phi^{J}\Phi^{K}/H\, ,
\end{equation}

\noindent
and using the condition Eq.~(\ref{eq:CIJKinvariance}) we find a linear
equation for the functions $L_{I}$:

\begin{equation}
\breve{\mathfrak{D}}^{2}L_{I} 
-g^{2} f_{IJ}{}^{L}f_{KL}{}^{M}\Phi^{J}\Phi^{K}L_{M}
=
0\, .   
\end{equation}

Finally, let us consider Eq.~(\ref{eq:omegaequation}). Defining $\hat{\omega}$
as

\begin{equation}
\hat{\omega} = \omega_{5}(dz+\chi) +\omega\, ,
\,\,\,\,\,\,
\mbox{where}
\,\,\,\,\,\,
\omega= \omega_{\underline{r}}dx^{r}\, ,  
\end{equation}

\noindent
Eq.~(\ref{eq:omegaequation}) gives an equation for $\omega_{5}$ whose general
solution is

\begin{equation}
\omega_{5}
=
M+16\sqrt{2} H^{-2} C_{IJK} \Phi^{I} \Phi^{J} \Phi^{K}
+3\sqrt{2} H^{-1} L_I \Phi^{I} \, ,
\,\,\,\,\,
\mbox{where}
\,\,\,\,\,
d\star_{3} d M=0\, ,
\end{equation}

\noindent
and the following equation for $\omega$:

\begin{equation}
\label{eq:omegad3}
\star_{3} d \omega
=
H dM-MdH
+3\sqrt{2} \left( \Phi^{I} \breve{\mathfrak{D}}L_{I}
-L_{I}\breve{\mathfrak{D}} \Phi^{I} \right)\, ,
\end{equation} 
 
\noindent
whose integrability condition $d^{2}\omega=0$ is satisfied wherever the above
equations for $H,M,\Phi^{I},L_{I}$ are satisfied.

Summarizing: we have identified a set of $z$-independent functions
$M,H,\Phi^{I},L_{I}$ and 1-forms $\omega,A^{I},\chi$ in $\mathbb{E}^{3}$ in
terms of which we can write all the building blocks of the 5-dimensional
timelike supersymmetric solutions admitting an isometry as follows:

\begin{eqnarray}
h_{I}/\hat{f} 
& = &
L_{I}+8C_{IJK}\Phi^{J}\Phi^{K}/H\, ,
\\
& & \nonumber \\
\hat{\omega} 
& = & 
\omega_{5}(dz+\chi) +\omega\, ,
\\
& & \nonumber \\
\label{eq:omega5}
\omega_{5}
& = &
M+16\sqrt{2} H^{-2} C_{IJK} \Phi^{I} \Phi^{J} \Phi^{K}
+3\sqrt{2} H^{-1} L_I \Phi^{I} \, ,
\\
& & \nonumber \\
\label{eq:Kromheimer}
\hat{A}^{I}
& = &
2\sqrt{6} \left[H^{-1}\Phi^{I} (dz+\chi)-\breve{A}^{I} \right]\, ,
\\
& & \nonumber \\
\hat{F}^{I}
& = &
2\sqrt{6} H^{-1} 
\left[\breve{\mathfrak{D}} \Phi^{I} \wedge (dz+\chi)
-\star_{3} H \breve{\mathfrak{D}} \Phi^{I} \right] \, ,
\end{eqnarray}

\noindent
provided that they satisfy the following set of equations:

\begin{eqnarray}
\label{eq:Mequation}
d\star_{3} d M 
& = &
0\, ,
\\
& & \nonumber \\
\label{eq:Hequation}
\star_{3}dH -d\chi  
& = &
0\, ,
\\
& & \nonumber \\
\label{eq:PhiIequation}
\star_{3}\breve{\mathfrak{D}} \Phi^{I}
- \breve{F}^{I}
& = &
0\, ,
\\
& & \nonumber \\
\label{eq:LIequation}
\breve{\mathfrak{D}}^{2}L_{I} 
-g^{2} f_{IJ}{}^{L}f_{KL}{}^{M}\Phi^{J}\Phi^{K}L_{M}
& = &
0\, ,  
\\
& & \nonumber \\
\label{eq:omegaequation2}
\star_{3} d \omega
-
\left\{
H dM-MdH
+3\sqrt{2} ( \Phi^{I} \breve{\mathfrak{D}}L_{I}
-L_{I}\breve{\mathfrak{D}} \Phi^{I} )
\right\}
& =&
0\, .
\end{eqnarray}

For symmetric real special manifolds we can use Eq.~(\ref{eq:f3symmetric}) to
write the metric function $\hat{f}$ explicitly in terms of the tensor $C_{IJK}$ and
the functions $M,H,\Phi^{I},L_{I}$:

\begin{equation}
\label{eq:f3symmetric-2}
  \begin{array}{rcl}
\hat{f}^{-3}
& = &
3^{3} C^{IJK}L_{I}L_{J}L_{K}
+3^{4}\cdot 2^{3}  C^{IJK}C_{KLM}L_{I}L_{J}\Phi^{L}\Phi^{M}/H
\\
& & \\
& &
+3\cdot 2^{6}L_{I}\Phi^{I}C_{JKL}\Phi^{J}\Phi^{K}\Phi^{L}/H^{2}
+2^{9}\left(C_{IJK}\Phi^{I}\Phi^{J}\Phi^{K}\right)^{2}/H^{3}\, .
\end{array}
\end{equation}

Let us compare the above formulae with those of the ungauged case (in
Ref.~\cite{Bellorin:2006yr} in our conventions). It is easy to see that all
the functions $M,H,\Phi^{I},L_{I}$ become standard harmonic functions in
$\mathbb{E}^{3}$. Furthermore, the functions $\Phi^{I}$ are related to the
functions $K^{I}$ used in that reference by

\begin{equation}
\Phi^{I} 
=
+\tfrac{1}{2\sqrt{2}}K^{I}\, .  
\end{equation}

%%%%%%%%%%%%%%%%%%%%%%%%%%%%%%%%%%%%%%%%%%%%%%%%%%%%%%%%%%%%%%%%%%%%%%
%%%%%%%%%%%%%%%%%%%%%%%%%%%%%%%%%%%%%%%%%%%%%%%%%%%%%%%%%%%%%%%%%%%%%%
%%%%%%%%%%%%%%%%%%%%%%%%%%%%%%%%%%%%%%%%%%%%%%%%%%%%%%%%%%%%%%%%%%%%%%
%%%%%%%%%%%%%%%%%%%%%%%%%%%%%%%%%%%%%%%%%%%%%%%%%%%%%%%%%%%%%%%%%%%%%%
\subsubsection{Dimensional reduction of the timelike supersymmetric solutions with one isometry}
%%%%%%%%%%%%%%%%%%%%%%%%%%%%%%%%%%%%%%%%%%%%%%%%%%%%%%%%%%%%%%%%%%%%%%
%%%%%%%%%%%%%%%%%%%%%%%%%%%%%%%%%%%%%%%%%%%%%%%%%%%%%%%%%%%%%%%%%%%%%%
%%%%%%%%%%%%%%%%%%%%%%%%%%%%%%%%%%%%%%%%%%%%%%%%%%%%%%%%%%%%%%%%%%%%%%
%%%%%%%%%%%%%%%%%%%%%%%%%%%%%%%%%%%%%%%%%%%%%%%%%%%%%%%%%%%%%%%%%%%%%%

The supersymmetric solutions that admit an additional isometry can be
dimensionally reduced to supersymmetric solutions of $\mathcal{N}=2,d=4$
supergravity using the formulae in Appendix~\ref{ap-reduction}\footnote{These
  formulae are valid for any field configuration, supersymmetric or
  not.}. Performing explicitly this reduction will allow us to simplify the
tasks of oxidation and reduction of supersymmetric solutions.

First of all, the metric of the 4-dimensional solutions obtained through the
dimensional reduction takes the conventional conformastationary form of the
timelike supersymmetric solutions of the $\mathcal{N}=2,d=4$ theory

\begin{equation}
\label{eq:conformastationarymetric}
ds^{2} 
= 
e^{2U}(dt+\omega)^{2} - e^{-2U}dx^{r}dx^{r}\, ,
\end{equation}

\noindent
where the 1-form $\omega= \omega_{\underline{r}}dx^{r}$ is precisely the
1-form given in Eq.~(\ref{eq:omegad3}) and the metric function $e^{-2U}$ is
given by

\begin{equation}
\label{eq:relationbetweenmetricfunctions}
e^{-2U}
=
2\sqrt{
\frac{
(\hat{f}^{\, -1}H)^{3}-(\omega_{5}H^{2})^{2}
}{4H^{2}}}\, .
\end{equation}

We can compare the equations satisfied by the building blocks of the timelike
supersymmetric solutions of gauged $\mathcal{N}=1,d=5$ supergravity
(\ref{eq:Mequation})-(\ref{eq:omegaequation2}) with the equations satisfied by
the building blocks of the timelike supersymmetric solutions of gauged
$\mathcal{N}=2,d=4$ supergravity Ref.~\cite{Hubscher:2008yz,Meessen:2012sr},
which we rewrite here for convenience adapting slightly the notation to avoid
confusion with the different accents used to distinguish the different gauge
fields:

\begin{eqnarray}
\label{eq:d4eqs-1}
-\tfrac{1}{\sqrt{2}}\star_{3}\breve{\mathfrak{D}} \mathcal{I}^{\Lambda}
-\breve{F}^{\Lambda}
& = &
0\, ,
\\
& & \nonumber \\
\label{eq:d4eqs-2}
\breve{\mathfrak{D}}^{2}\mathcal{I}_{\Lambda} 
-\tfrac{1}{2}g^{2} f_{\Lambda\Sigma}{}^{\Omega}f_{\Delta\Omega}{}^{\Gamma}
\mathcal{I}^{\Sigma}\mathcal{I}^{\Delta}\mathcal{I}_{\Gamma}
& = &
0\, ,  
\\
& & \nonumber \\
\label{eq:d4eqs-3}
\star_{3}d\omega
-
2\left[
\mathcal{I}_{\Lambda}\breve{\mathfrak{D}}\mathcal{I}^{\Lambda}
-
\mathcal{I}^{\Lambda}\breve{\mathfrak{D}}\mathcal{I}_{\Lambda}
\right]
& = &
0
\, ,  
\end{eqnarray}

\noindent
where $\breve{\mathfrak{D}}$ is the gauge covariant derivative associated to
the modified gauge connection in $\mathbb{E}^{3}$

\begin{equation}
\breve{A}^{\Lambda}{}_{\underline{m}}
\equiv
A^{\Lambda}{}_{\underline{m}}
-\omega_{\underline{m}}
A^{\Lambda}{}_{t}\, .
\end{equation}

The notation that we are using has implicit the identification of the gauge
potentials $\breve{A}$ coming from 5 and 4 dimensions, except for $\Lambda=0$.
Using the formulae in Appendix~\ref{ap-reduction} with the modifications
explained in the last paragraph we can identify\footnote{The $0$th components
  are never gauged if the dimensional reduction is simple (not
  generalized).}

\begin{equation}
\chi_{\underline{m}} = -2\sqrt{2}\breve{A}^{0}{}_{\underline{m}}\, ,  
\end{equation}

\noindent
which leads to the identifications 

\begin{equation}
\label{eq:identifications}
\Phi^{I} = -\tfrac{1}{\sqrt{2}} \mathcal{I}^{I+1}\, ,
\hspace{1cm}
L_{I} = \tfrac{2}{3}\mathcal{I}_{I+1}\, ,  
\hspace{1cm}
H = 2\mathcal{I}^{0}\, ,
\hspace{1cm}
M = -\mathcal{I}_{0}\, .
\end{equation}

These are the only formulae we need to relate timelike supersymmetric
solutions in $\mathcal{N}=1,d=5$ supergravity with one additional isometry to
timelike supersymmetric solutions in cubic model of $\mathcal{N}=2,d=4$
supergravity with $\mathcal{I}^{0}\neq 0$\footnote{Those with
  $\mathcal{I}^{0}= 0$ are related to null supersymmetric 5-dimensional
  solutions.}.

For symmetric real special scalar manifolds we can use the explicit form of
$\hat{f}$ in Eq.~(\ref{eq:f3symmetric-2}) together with the expression for
$\omega_{5}$ in Eq.~(\ref{eq:omega5}) to get

\begin{equation}
  \begin{array}{rcl}
e^{-2U}
& = &
2\left\{
\tfrac{3^{3}}{4}HC^{IJK}L_{I}L_{J}L_{K}
-2^{7/2}M C_{IJK}\Phi^{I}\Phi^{J}\Phi^{K} 
+2\cdot 3^{4}C^{IJK}C_{KLM}L_{I}L_{J}\Phi^{L}\Phi^{M}
\right.
\\
& & \\
& & 
\left.
-\tfrac{3^{2}}{2} \left(L_{I}\Phi^{I} \right)^{2}
-\tfrac{3}{\sqrt{2}} HM L_{I}\Phi^{I} 
-\tfrac{1}{4}M^{2}H^{2}
\right\}^{1/2}\, .  
\end{array}
\end{equation}

\noindent
Then, using the identifications Eqs.~(\ref{eq:identifications}) together with
the second of Eqs.~(\ref{eq:KKreductionN1d5N2d4}) we get

\begin{equation}
  \begin{array}{rcl}
e^{-2U}
& = &
2\left\{
(d^{ijk}\mathcal{I}_{j}\mathcal{I}_{l}
-\tfrac{2}{3}\mathcal{I}_{0}\mathcal{I}^{i})
(d_{ilm}\mathcal{I}^{l}\mathcal{I}^{m}
+\tfrac{2}{3}\mathcal{I}^{0}\mathcal{I}_{i})
+\tfrac{4}{9}\mathcal{I}^{0}\mathcal{I}_{0}\mathcal{I}^{i}\mathcal{I}_{i}
\right.
\\
& & \\
& & 
\left.
-(\mathcal{I}^{0}\mathcal{I}_{0}+\mathcal{I}^{i}\mathcal{I}_{i})^{2}
\right\}^{1/2}\, .  
\end{array}
\end{equation}

%%%%%%%%%%%%%%%%%%%%%%%%%%%%%%%%%%%%%%%%%%%%%%%%%%%%%%%%%%%%%%%%%%%%%%
%%%%%%%%%%%%%%%%%%%%%%%%%%%%%%%%%%%%%%%%%%%%%%%%%%%%%%%%%%%%%%%%%%%%%%
%%%%%%%%%%%%%%%%%%%%%%%%%%%%%%%%%%%%%%%%%%%%%%%%%%%%%%%%%%%%%%%%%%%%%%
%%%%%%%%%%%%%%%%%%%%%%%%%%%%%%%%%%%%%%%%%%%%%%%%%%%%%%%%%%%%%%%%%%%%%%
\subsection{Null supersymmetric solutions}
%%%%%%%%%%%%%%%%%%%%%%%%%%%%%%%%%%%%%%%%%%%%%%%%%%%%%%%%%%%%%%%%%%%%%%
%%%%%%%%%%%%%%%%%%%%%%%%%%%%%%%%%%%%%%%%%%%%%%%%%%%%%%%%%%%%%%%%%%%%%%
%%%%%%%%%%%%%%%%%%%%%%%%%%%%%%%%%%%%%%%%%%%%%%%%%%%%%%%%%%%%%%%%%%%%%%
%%%%%%%%%%%%%%%%%%%%%%%%%%%%%%%%%%%%%%%%%%%%%%%%%%%%%%%%%%%%%%%%%%%%%%

The general form of the null supersymmetric solutions of $\mathcal{N}=1,d=5$
SEYM is quite involved \cite{Bellorin:2007yp}, but it simplifies dramatically
when one assumes the existence of an additional isometry so that all the
fields are independent of the two null coordinates $u$ and $v$. These are the
solutions which will become timelike supersymmetric solutions of
$\mathcal{N}=2,d=4$ SEYM upon dimensional reduction and, therefore, we are
going to describe only these.

%%%%%%%%%%%%%%%%%%%%%%%%%%%%%%%%%%%%%%%%%%%%%%%%%%%%%%%%%%%%%%%%%%%%%%
%%%%%%%%%%%%%%%%%%%%%%%%%%%%%%%%%%%%%%%%%%%%%%%%%%%%%%%%%%%%%%%%%%%%%%
%%%%%%%%%%%%%%%%%%%%%%%%%%%%%%%%%%%%%%%%%%%%%%%%%%%%%%%%%%%%%%%%%%%%%%
%%%%%%%%%%%%%%%%%%%%%%%%%%%%%%%%%%%%%%%%%%%%%%%%%%%%%%%%%%%%%%%%%%%%%%
\subsubsection{$u$-independent null supersymmetric solutions}
%%%%%%%%%%%%%%%%%%%%%%%%%%%%%%%%%%%%%%%%%%%%%%%%%%%%%%%%%%%%%%%%%%%%%%
%%%%%%%%%%%%%%%%%%%%%%%%%%%%%%%%%%%%%%%%%%%%%%%%%%%%%%%%%%%%%%%%%%%%%%
%%%%%%%%%%%%%%%%%%%%%%%%%%%%%%%%%%%%%%%%%%%%%%%%%%%%%%%%%%%%%%%%%%%%%%
%%%%%%%%%%%%%%%%%%%%%%%%%%%%%%%%%%%%%%%%%%%%%%%%%%%%%%%%%%%%%%%%%%%%%%

The metric of the general null supersymmetric solutions of $\mathcal{N}=1,d=5$
SEYM can always brought into the form \cite{Bellorin:2007yp}\footnote{We have
  changed the notation and normalization with respect to \cite{Bellorin:2007yp} to avoid possible confusions between
  the objects that appear in the null and timelike cases.}

%\omega \rightarrow \sqrt{2} \omega
%H      \rightarrow K
%f      \rightarrow \ell

\begin{equation}
\label{eq:nullmetric}
ds^{2}= 2\ell du(dv+K du+\sqrt{2} \omega)-\ell^{-2}dx^{r}dx^{r}\, ,  
\end{equation}

\noindent
where the functions $\ell,K$ and the 1-form
$\omega=\omega_{\underline{r}}dx^{r}$ are $v$-independent. We are going to
assume also $u$-independence of all the fields throughout.

After the partial gauge fixing $A^{I}{}_{\underline{v}}=0$, the gauge fields
are decomposed as\footnote{As the notation suggests, the gauge fields
  $\breve{A}^{I}$ are the same as the $\mathcal{N}=2,d=4$ fields denoted with
  the same symbols, according to the general formulae of
  Appendix~\ref{ap-reduction}. The same is true of the 1-form $\omega$.}

\begin{equation}
\label{eq:potentialdecomposition}
A^{I} = A^{I}{}_{\underline{u}} du -2\sqrt{6} \breve{A}^{I} \, ,
\hspace{1cm}
\breve{A}^{I} = \breve{A}^{I}{}_{\underline{r}}dx^{r}\, ,
\end{equation}

\noindent
and the vector field strengths take the form\footnote{All the operators in the
  r.h.s.~are defined in $\mathbb{E}^{3}$.}

\begin{equation}
\label{eq:vectorfieldstrengths}
F^{I} 
= 
(\sqrt{2/3} \ell^{2}h^{I}\star_{3}d\omega-\psi^{I})\wedge du
+\sqrt{3}\star_{3} \breve{\mathfrak{D}}(h^{I}/\ell)\, ,
\end{equation}

\noindent
where the $\psi^{I}$ are some 1-forms in $\mathbb{E}^{3}$ satisfying

\begin{equation}
\label{eq:psiconstraint}
h_{I}\psi^{I}=0\, ,  
\end{equation}

\noindent
to be determined and $\breve{\mathfrak{D}}$ is the gauge-covariant derivative
on $\mathbb{E}^{3}$ with respect to the connection $\breve{A}^{I}$.

Finally, the scalar fields will be determined by the equations obeyed by the
scalar functions $h^{I}$, which follow from the equations of
motion.\footnote{The field configurations that we have just described are
  automatically supersymmetric, but not necessarily solutions of all the
  equations of motion and Bianchi identities \cite{Bellorin:2007yp}.}

Let us start by analyzing the Bianchi identities of the vector field
strength. They lead to the following two sets of equations:

\begin{eqnarray}
\label{eq:nullbianchi1}
-\tfrac{1}{2\sqrt{2}}\star_{3} 
\breve{\mathfrak{D}}(h^{I}/\ell)
-\breve{F}^{I} 
& = & 
0\, , 
\\ 
& & \nonumber \\ 
\label{eq:nullbianchi2}
\breve{\mathfrak{D}} A^{I}{}_{\underline{u}} 
%-\partial_{\underline{u}} \breve{A}^{I}
-\sqrt{2/3} \ell^{2}h^{I}\star_{3}d\omega +\psi^{I}
& = &
0\, .
`\end{eqnarray}

\noindent
Eq.~(\ref{eq:nullbianchi1}) is the Bogomol'nyi equation on $\mathbb{E}^{3}$
and, thus, we define the Higgs field

\begin{equation}
\Sigma^{I} \equiv -\tfrac{1}{2\sqrt{2}} h^{I}/\ell\, .
\end{equation}

Multiplying Eq.~(\ref{eq:nullbianchi2}) by $h_{I}$ and using
Eq.~(\ref{eq:psiconstraint}) together with $h_{I}h^{I}=1$ we get the equation that
defines $\omega$

\begin{equation}
d\omega 
=
\sqrt{3/2}\ell^{-2}
\star_{3}\left\{
h_{I}\breve{\mathfrak{D}} A^{I}{}_{\underline{u}} 
%-h_{I}\partial_{\underline{u}} \breve{A}^{I}
\right\}\, .
\end{equation}

\noindent
Defining the functions

\begin{equation}
\label{eq:KsubIdef}
K_{I} \equiv C_{IJK}\Sigma^{J}A^{K}{}_{\underline{u}}\, ,  
\end{equation}

\noindent
the above equation takes a much more familiar form

\begin{equation}
d\omega 
=
4\sqrt{6}
\star_{3}\left\{
\Sigma^{I}\breve{\mathfrak{D}}K_{I}
-
K_{I}\breve{\mathfrak{D}} \Sigma^{I}
%-C_{IJK}\Sigma^{I}\Sigma^{J}\partial_{\underline{u}} \breve{A}^{K}
\right\}\, ,
\end{equation}

\noindent
whose integrability condition is

\begin{equation}
\label{eq:omegaequationintegrabilityconditionnullcase}
\Sigma^{I}\breve{\mathfrak{D}}^{2}K_{I} =0\, .
\end{equation}

Given the functions $\Sigma^{I},K_{I}$ and the gauge fields $\breve{A}^{I}$ we
can solve this equation for $\omega$. It should be possible to find the
functions $A^{I}{}_{\underline{u}}$ in terms of
$\Sigma^{I},K_{I}$\footnote{This will certainly be the case for the particular
  model we are going to study, but we have not found (even for just the
  symmetric case) a general way of solving Eq.~(\ref{eq:KsubIdef}) for
  $A^{I}{}_{\underline{u}}$.} and, plugging these result in
Eq.~(\ref{eq:nullbianchi2}), compute directly the 1-forms $\psi^{I}$.

From the Maxwell equations one obtains the equations that determine the
functions $K_{I}$:

\begin{equation}
\label{eq:KsubIequation}
\breve{\mathfrak{D}}^{2}K_{I}
-g^{2}f_{IJ}{}^{L}f_{KL}{}^{M}\Sigma^{J}\Sigma^{K}K_{M} =0\, ,  
\end{equation}

\noindent
from which the integrability condition
Eq.~(\ref{eq:omegaequationintegrabilityconditionnullcase}) follows
automatically.

Finally, defining

\begin{equation}
N\equiv K -\sqrt{2}A^{I}{}_{\underline{u}}K_{I}\, ,  
\end{equation}

\noindent
the last non-trivial equation of motion, from the Einstein equations, takes
the simple form 

\begin{equation}
\nabla^{2}N=0\, .  
\end{equation}

Summarizing: we have identified a set of $u$-independent functions
$\Sigma^{I},K_{I},N$ and 1-forms $\omega,\breve{A}^{I}$ on $\mathbb{E}^{3}$ in
terms of which we can write all the building blocks of the 5-dimensional
$u$-independent null supersymmetric solutions, assuming we can solve
Eq.~(\ref{eq:KsubIdef}) for $A^{I}{}_{\underline{u}}$, as follows:

\begin{eqnarray}
h^{I}/\ell
& = &
-2\sqrt{2}\Sigma^{I}\, ,
\\
& & \nonumber \\   
K & = &
N+\sqrt{2}A^{I}{}_{\underline{u}}K_{I}\, ,
\\
& & \nonumber \\   
A^{I} 
& = &
A^{I}{}_{\underline{u}} du +2\sqrt{6} \breve{A}^{I} \, ,
\\
& & \nonumber \\
F^{I} 
& = & 
\breve{\mathfrak{D}} A^{I}{}_{\underline{u}}\wedge du
+\sqrt{3}\star_{3} \breve{\mathfrak{D}}(h^{I}/\ell)\, ,
\end{eqnarray}

\noindent
provided the following equations are satisfied\footnote{The gauge coupling
  constant is the 4-dimensional one.}:

\begin{eqnarray}
\label{eq:SigmaIequation}
\star_{3} 
\breve{\mathfrak{D}}\Sigma^{I}
-\breve{F}^{I} 
& = & 
0\, , 
\\
& & \nonumber \\
\breve{\mathfrak{D}}^{2}K_{I}
-g^{2}f_{IJ}{}^{L}f_{KL}{}^{M}\Sigma^{J}\Sigma^{K}K_{M} 
& = &
0\, ,  
\\
& & \nonumber \\
d\omega 
-
4\sqrt{6}
\star_{3}\left\{
\Sigma^{I}\breve{\mathfrak{D}}K_{I}
-
K_{I}\breve{\mathfrak{D}} \Sigma^{I}
%-C_{IJK}\Sigma^{I}\Sigma^{J}\partial_{\underline{u}} \breve{A}^{K}
\right\}
& = &
0\, ,
\\
& & \nonumber \\
\label{eq:Nequation}
\nabla^{2}N
& = & 0\, .
\end{eqnarray}

Using Eq.~(\ref{eq:Ch3=1}), we find a general expression for $\ell$:

\begin{equation}
\ell^{-3} = -2^{9/2}C_{IJK}\Sigma^{I}\Sigma^{J}\Sigma^{K}\, .     
\end{equation}

%%%%%%%%%%%%%%%%%%%%%%%%%%%%%%%%%%%%%%%%%%%%%%%%%%%%%%%%%%%%%%%%%%%%%%
%%%%%%%%%%%%%%%%%%%%%%%%%%%%%%%%%%%%%%%%%%%%%%%%%%%%%%%%%%%%%%%%%%%%%%
%%%%%%%%%%%%%%%%%%%%%%%%%%%%%%%%%%%%%%%%%%%%%%%%%%%%%%%%%%%%%%%%%%%%%%
%%%%%%%%%%%%%%%%%%%%%%%%%%%%%%%%%%%%%%%%%%%%%%%%%%%%%%%%%%%%%%%%%%%%%%
\subsubsection{Dimensional reduction of the $u$-independent null
  supersymmetric solutions}
%%%%%%%%%%%%%%%%%%%%%%%%%%%%%%%%%%%%%%%%%%%%%%%%%%%%%%%%%%%%%%%%%%%%%%
%%%%%%%%%%%%%%%%%%%%%%%%%%%%%%%%%%%%%%%%%%%%%%%%%%%%%%%%%%%%%%%%%%%%%%
%%%%%%%%%%%%%%%%%%%%%%%%%%%%%%%%%%%%%%%%%%%%%%%%%%%%%%%%%%%%%%%%%%%%%%
%%%%%%%%%%%%%%%%%%%%%%%%%%%%%%%%%%%%%%%%%%%%%%%%%%%%%%%%%%%%%%%%%%%%%%

Using the general formulae in Appendix~\ref{ap-reduction}, the $u$-independent
solutions that we have considered can be dimensionally reduced to timelike
supersymmetric solutions of $\mathcal{N}=2,d=4$ SEYM along the spacelike
coordinate $z$ defined by

\begin{equation}
u =\tfrac{1}{\sqrt{2}}(t+z)\, ,
\hspace{1cm}  
v =\tfrac{1}{\sqrt{2}}(t-z)\, ,
\end{equation}

\noindent
with metrics of the form Eq.~(\ref{eq:conformastationarymetric}) where the
1-form $\omega= \omega_{\underline{r}}dx^{r}$ is precisely the 1-form given in
Eq.~(\ref{eq:nullmetric}) and the metric function $e^{-2U}$ is given by

\begin{equation}
\label{eq:relationbetweenmetricfunctionsnullcase}
e^{-2U}
=
\sqrt{
\ell^{-3}(1-K)
}
=
\sqrt{
-2^{9/2}C_{IJK}\Sigma^{I}\Sigma^{J}\Sigma^{K}
(1-N-\sqrt{2}A^{I}{}_{\underline{u}}K_{I})
}
\, .
\end{equation}

In order to express entirely the metric function in terms of the functions
$K_{I},\Sigma^{I},N$ we need to solve Eq.~(\ref{eq:KsubIdef}) for
$A^{I}{}_{\underline{u}}$ as a function of $K_{I},\Sigma^{I}$, which we do not
know how to do in general. We can still compare the equations satisfied by
these functions (\ref{eq:SigmaIequation})-(\ref{eq:Nequation}) with those
satisfied by $\mathcal{I}^{\Lambda},\mathcal{I}_{\Lambda}$ in
$\mathcal{N}=2,d=4$ SEYM (\ref{eq:d4eqs-1})-(\ref{eq:d4eqs-3}) knowing that
the vector fields $\breve{A}^{I}$ and the 1-form $\omega$ are the same
objects. We find that 

\begin{equation}
\Sigma^{I}=-\tfrac{1}{\sqrt{2}}\mathcal{I}^{I+1}\, ,
\hspace{1cm}  
K_{I}=-\tfrac{1}{2\sqrt{3}}\mathcal{I}_{I+1}\, ,
\end{equation}

\noindent
while $N$ must be proportional to either $\mathcal{I}^{0}$ or
$\mathcal{I}_{0}$. Since a wave moving in the internal $z$ direction should
give rise to a 4-dimensional electric charge, it must be

\begin{equation}
N\sim \mathcal{I}_{0}\, ,  
\end{equation}

\noindent
but the precise coefficient cannot be determined from this comparison
alone. We have to find a more explicit expression for $e^{-2U}$.

%%%%%%%%%%%%%%%%%%%%%%%%%%%%%%%%%%%%%%%%%%%%%%%%%%%%%%%%%%%%%%%%%%%%%%
%%%%%%%%%%%%%%%%%%%%%%%%%%%%%%%%%%%%%%%%%%%%%%%%%%%%%%%%%%%%%%%%%%%%%%
%%%%%%%%%%%%%%%%%%%%%%%%%%%%%%%%%%%%%%%%%%%%%%%%%%%%%%%%%%%%%%%%%%%%%%
%%%%%%%%%%%%%%%%%%%%%%%%%%%%%%%%%%%%%%%%%%%%%%%%%%%%%%%%%%%%%%%%%%%%%%
\section{5-dimensional supersymmetric non-Abelian solutions of the
  SU$(2)$-gauged ST$[2,5]$ model}
\label{sec-solutions}
%%%%%%%%%%%%%%%%%%%%%%%%%%%%%%%%%%%%%%%%%%%%%%%%%%%%%%%%%%%%%%%%%%%%%%
%%%%%%%%%%%%%%%%%%%%%%%%%%%%%%%%%%%%%%%%%%%%%%%%%%%%%%%%%%%%%%%%%%%%%%
%%%%%%%%%%%%%%%%%%%%%%%%%%%%%%%%%%%%%%%%%%%%%%%%%%%%%%%%%%%%%%%%%%%%%%
%%%%%%%%%%%%%%%%%%%%%%%%%%%%%%%%%%%%%%%%%%%%%%%%%%%%%%%%%%%%%%%%%%%%%%

In this section we are going to consider a particular model of
$\mathcal{N}=1,d=5$ supergravity that admits an SU$(2)$ gauging.  This model
is related to the SU$(2)$-gauged ST$[2,5]$ model of $\mathcal{N}=2,d=4$
supergravity some of whose solutions we have studied in
Ref.~\cite{Bueno:2014mea}. We will use the relations derived in the previous
section to find relations between the non-Abelian supersymmetric solutions of
both theories.

We start by describing the 4- and 5-dimensional models and their SU$(2)$
gauging.

%%%%%%%%%%%%%%%%%%%%%%%%%%%%%%%%%%%%%%%%%%%%%%%%%%%%%%%%%%%%%%%%%%%%%%
%%%%%%%%%%%%%%%%%%%%%%%%%%%%%%%%%%%%%%%%%%%%%%%%%%%%%%%%%%%%%%%%%%%%%%
%%%%%%%%%%%%%%%%%%%%%%%%%%%%%%%%%%%%%%%%%%%%%%%%%%%%%%%%%%%%%%%%%%%%%%
%%%%%%%%%%%%%%%%%%%%%%%%%%%%%%%%%%%%%%%%%%%%%%%%%%%%%%%%%%%%%%%%%%%%%%
\subsection{The models}
%%%%%%%%%%%%%%%%%%%%%%%%%%%%%%%%%%%%%%%%%%%%%%%%%%%%%%%%%%%%%%%%%%%%%%
%%%%%%%%%%%%%%%%%%%%%%%%%%%%%%%%%%%%%%%%%%%%%%%%%%%%%%%%%%%%%%%%%%%%%%
%%%%%%%%%%%%%%%%%%%%%%%%%%%%%%%%%%%%%%%%%%%%%%%%%%%%%%%%%%%%%%%%%%%%%%
%%%%%%%%%%%%%%%%%%%%%%%%%%%%%%%%%%%%%%%%%%%%%%%%%%%%%%%%%%%%%%%%%%%%%%

The ST$[2,5]$ model is a cubic model of $\mathcal{N}=2,d=4$ supergravity
coupled to 5 vector multiplets \textit{i.e.}~a model with a prepotential of
the form

\begin{equation}
\label{eq:prepotential}
\mathcal{F}  
= 
-\tfrac{1}{3!}
\frac{d_{ijk}\mathcal{X}^{i}\mathcal{X}^{j}\mathcal{X}^{k}}{\mathcal{X}^{0}}\, ,
\,\,\,\,\,\,
i=1,2\cdots,5
\end{equation}

\noindent
where the fully symmetric tensor $d_{ijk}$ has as only non-vanishing 
components

\begin{equation}
d_{1\alpha\beta}= \eta_{\alpha\beta}\, ,
\,\,\,\,\,
\mbox{where}
\,\,\,\,\,
(\eta_{\alpha\beta}) = \mathrm{diag}(+-\dotsm -)\, ,
\,\,\,\,\,
\mbox{and}
\,\,\,\,\,
\alpha,\beta=2,\cdots,5\, .
\end{equation}

\noindent
The 5 complex scalars parametrize the coset space 

\begin{equation}
\frac{\mathrm{SL}(2,\mathbb{R})}{\mathrm{SO}(2)}
\times
\frac{\mathrm{SO}(2,4)}{\mathrm{SO}(2)\times \mathrm{SO}(4)}\, , 
\end{equation}

\noindent
and the group SO$(3)$ acts in the adjoint on the coordinates
$\alpha=3,4,5$. These are the directions we are going to gauge and we will
denote them with capital $A,B,\ldots$. This is the only information we need in
order to construct supersymmetric solutions, but more details on the
construction of this theory can be found in Ref.~\cite{Bueno:2014mea}. We will
need the form of the metric function in terms of the functions
$\mathcal{I}^{M}$:

\begin{equation}
\label{eq:d4metricfunctionST24}
e^{-2U}
=
2\sqrt{
(\mathcal{I}^{\alpha}\mathcal{I}^{\beta}\eta_{\alpha\beta}
+2\mathcal{I}^{0}\mathcal{I}_{1})
(\mathcal{I}_{\alpha}\mathcal{I}_{\beta}\eta^{\alpha\beta}
-2\mathcal{I}^{1}\mathcal{I}_{0})
-(\mathcal{I}^{0}\mathcal{I}_{0}-\mathcal{I}^{1}\mathcal{I}_{1}
+\mathcal{I}^{\alpha}\mathcal{I}_{\alpha})^{2}\, .
}\, ,
\end{equation}

The models of the ST$[2,n]$ family are related to the effective theory of the
Heterotic string and compactified on $T^{6}$ by a consistent truncation: the
10-dimensional effective theory is $\mathcal{N}=1,d=10$ supergravity coupled
to $16$ 10-dimensional vector multiplets with gauge group U$(1)$. Upon
dimensional reduction on a generic $T^{6}$ one gets $\mathcal{N}=4,d=4$
supergravity coupled to $16+6=22$ vector multiplets, whose duality group is 

\begin{equation}
\frac{\mathrm{SL}(2,\mathbb{R})}{\mathrm{SO}(2)}
\times
\frac{\mathrm{SO}(6,22)}{\mathrm{SO}(6)\times \mathrm{SO}(22)}\, . 
\end{equation}

Observe that $\mathrm{SO}(6)$ acts on the $6$ vectors in the supergravity
multiplet and $\mathrm{SO}(22)$ on the $22$ matter vector fields.  The coset
$\mathrm{SL}(2,\mathbb{R})/\mathrm{SO}(2)$ is parametrized by the only scalar
in the supergravity multiplet. A consistent truncation to $\mathcal{N}=2,d=4$
eliminates 4 vectors from the $\mathcal{N}=4$ supergravity multiplet and one
of the remaining two vectors becomes a matter vector field from the
$\mathcal{N}=2$ point of view and comes in the same multiplet as the complex
scalar that parametrizes the coset space
$\mathrm{SL}(2,\mathbb{R})/\mathrm{SO}(2)$. The result is a ST$[2,23]$ model
from which one can consistently eliminate vector multiplets to arrive to the
ST$[2,5]$ model we are dealing with.

This is the story at a generic point in the moduli space of the Heterotic
strings on $T^{6}$. At certain points, though, there is a enhancement of gauge
symmetry usually associated to an increase in the number of massless vector
fields that we must take into account in the effective theory.  Our
SU$(2)$-gauged model of $\mathcal{N}=2,d=4$ supergravity can be interpreted as
the effective theory describing the simplest of these situations in which the
enhancement of gauge symmetry arises in the sector of the $16$ original
10-dimensional vector fields.

The ST$[2,5]$ model is related to a model of $\mathcal{N}=1,d=5$ supergravity
coupled to 4 vector multiplets determined by the tensor $C_{i-1,j-1,k-1} =
\tfrac{1}{6}d_{ijk}$ so its only non-vanishing components are

\begin{equation}
C_{0xy}= \tfrac{1}{6}\eta_{xy}\, ,
\mbox{where}
\,\,\,\,\,
(\eta_{xy}) = \mathrm{diag}(+-\dotsm -)\, ,
\,\,\,\,\,
\mbox{and}
\,\,\,\,\,
x,y=1,\cdots,4\, .
\end{equation}

The 4 real scalars in the vector multiplets parametrize the coset space

\begin{equation}
\frac{\mathrm{SO}(1,4)}{\mathrm{SO}(4)}\, .
\end{equation}

\noindent
Now the group SO$(3)$ acts in the adjoint on the coordinates $x=2,3,4$ and, if
we gauge it, the theory goes to the gauged 4-dimensional model we just
discussed. It should be obvious after the 4-dimensional discussion that that
this model can be interpreted as a truncation of the effective theory of the
Heterotic string compactified on $T^{5}$.

Again, we do not need many more details of the theory in order to
construct supersymmetric solutions. For timelike supersymmetric solutions
admitting an additional isometry we will need the metric function, which
follows directly from the generic expression Eq.~(\ref{eq:f3symmetric-2})

\begin{equation}
\label{eq:d5metricfunctionST24}
  \begin{array}{rcl}
\hat{f}^{\, -1}
& = &
H^{-1}
\left\{
\tfrac{1}{4}
\left(
6HL_{0} +8\eta_{xy}\Phi^{x}\Phi^{y}
\right)
\left[
9H^{2}\eta^{xy}L_{x}L_{y} +48 H\Phi^{0}L_{x}\Phi^{x}
\right.
\right.
\\
& & \\
& & 
\left.\left.
+64(\Phi^{0})^{2}\eta_{xy}\Phi^{x}\Phi^{y}
\right]
\right\}^{1/3}
\end{array}
\end{equation}

This metric function and the 4-dimensional one $e^{-2U}$ are related by
Eq.~(\ref{eq:relationbetweenmetricfunctions}) using Eq.~(\ref{eq:omega5}) and
the relations between the functions $\mathcal{I}^{M}$ and $H,M,L_{I},\Phi^{I}$
in Eqs.~(\ref{eq:identifications}), which we rewrite for this specific pair of
models for convenience:

\begin{equation}
\label{eq:identificationsST24}
\begin{array}{rclrclrclrcl}
H & = & 2\mathcal{I}^{0}\, ,
\hspace{.3cm}
&
\Phi^{0} & = &-\tfrac{1}{\sqrt{2}} \mathcal{I}^{1}\, ,
\hspace{.3cm}
&
\Phi^{1} & = & -\tfrac{1}{\sqrt{2}} \mathcal{I}^{2}\, ,
\hspace{.3cm}
&
\Phi^{A} & = & -\tfrac{1}{\sqrt{2}} \mathcal{I}^{A}\, ,
\\
& & & & & & & & & & & \\
M & = & -\mathcal{I}_{0}\, ,
&
L_{0} & = &\tfrac{2}{3}\mathcal{I}_{1}\, ,  
&
L_{1} & = & \tfrac{2}{3}\mathcal{I}_{2}\, ,
&  
L_{A} & = & \tfrac{2}{3}\mathcal{I}_{A}\, ,  
\\
\end{array}
\end{equation}

For $u$-independent null supersymmetric solutions we first need to solve
Eq.~(\ref{eq:KsubIdef}) for $A^{I}{}_{\underline{u}}$. For this model, we find

\begin{equation}
A^{0}{}_{\underline{u}}
=
6\frac{\Sigma^{x}K_{x}-\Sigma^{0}K_{0}}{(\eta \Sigma\Sigma)}\, ,
\hspace{1cm}
A^{x}{}_{\underline{u}}
=
6\frac{\eta^{xy}K_{y}(\eta \Sigma\Sigma)
  -\Sigma^{x}(\Sigma^{y}K_{y}-\Sigma^{0}K_{0})}{\Sigma^{0}(\eta
  \Sigma\Sigma)}\, ,  
\end{equation}

\noindent
where $(\eta \Sigma\Sigma) \equiv \eta_{xy} \Sigma^{x}\Sigma^{y}$, so that

% \begin{equation}
% K
% =   
% \frac{N\Sigma^{0}(\eta
%   \Sigma\Sigma) +6\sqrt{2}(\eta LL)(\eta \Sigma\Sigma)
%   -6\sqrt{2}(\Sigma^{y}K_{y}-\Sigma^{0}K_{0})^{2}}{\Sigma^{0}(\eta
%   \Sigma\Sigma)}\, ,  
% \end{equation}

\begin{equation}
e^{-2U}
=
2\sqrt{
(\mathcal{I}^{\alpha}\mathcal{I}^{\beta}\eta_{\alpha\beta})
[\mathcal{I}_{\alpha}\mathcal{I}_{\beta}\eta^{\alpha\beta}
+\mathcal{I}^{1}(1-N)]
-(-\mathcal{I}^{1}\mathcal{I}_{1}
+\mathcal{I}^{\alpha}\mathcal{I}_{\alpha})^{2}\, .
}\, ,
\end{equation}

\noindent
and we arrive at the following identifications

\begin{equation}
\label{eq:identificationsST24null}
\begin{array}{rclrclrclrcl}
0 & = & \mathcal{I}^{0}\, ,
\hspace{.3cm}
&
\Sigma^{0} & = & -\tfrac{1}{\sqrt{2}} \mathcal{I}^{1}\, ,
\hspace{.3cm}
&
\Sigma^{1} & = & -\tfrac{1}{\sqrt{2}} \mathcal{I}^{2}\, ,
\hspace{.3cm}
&
\Sigma^{A} & = & -\tfrac{1}{\sqrt{2}} \mathcal{I}^{A}\, ,
\\
& & & & & & & & & & & \\
N & = & 1+2\mathcal{I}_{0}\, ,
&
K_{0} & = & -\tfrac{1}{2\sqrt{3}}\mathcal{I}_{1}\, ,  
&
K_{1} & = & -\tfrac{1}{2\sqrt{3}}\mathcal{I}_{2}\, ,
&  
K_{A} & = & -\tfrac{1}{2\sqrt{3}}\mathcal{I}_{A}\, .
\\
\end{array}
\end{equation}

%%%%%%%%%%%%%%%%%%%%%%%%%%%%%%%%%%%%%%%%%%%%%%%%%%%%%%%%%%%%%%%%%%%%%%
%%%%%%%%%%%%%%%%%%%%%%%%%%%%%%%%%%%%%%%%%%%%%%%%%%%%%%%%%%%%%%%%%%%%%%
%%%%%%%%%%%%%%%%%%%%%%%%%%%%%%%%%%%%%%%%%%%%%%%%%%%%%%%%%%%%%%%%%%%%%%
%%%%%%%%%%%%%%%%%%%%%%%%%%%%%%%%%%%%%%%%%%%%%%%%%%%%%%%%%%%%%%%%%%%%%%
\subsection{The  solutions}
%%%%%%%%%%%%%%%%%%%%%%%%%%%%%%%%%%%%%%%%%%%%%%%%%%%%%%%%%%%%%%%%%%%%%%
%%%%%%%%%%%%%%%%%%%%%%%%%%%%%%%%%%%%%%%%%%%%%%%%%%%%%%%%%%%%%%%%%%%%%%
%%%%%%%%%%%%%%%%%%%%%%%%%%%%%%%%%%%%%%%%%%%%%%%%%%%%%%%%%%%%%%%%%%%%%%
%%%%%%%%%%%%%%%%%%%%%%%%%%%%%%%%%%%%%%%%%%%%%%%%%%%%%%%%%%%%%%%%%%%%%%

We are ready to put to work the machinery developed in the previous
sections. We are going to consider the simplest cases first.

%%%%%%%%%%%%%%%%%%%%%%%%%%%%%%%%%%%%%%%%%%%%%%%%%%%%%%%%%%%%%%%%%%%%%%
%%%%%%%%%%%%%%%%%%%%%%%%%%%%%%%%%%%%%%%%%%%%%%%%%%%%%%%%%%%%%%%%%%%%%%
%%%%%%%%%%%%%%%%%%%%%%%%%%%%%%%%%%%%%%%%%%%%%%%%%%%%%%%%%%%%%%%%%%%%%%
%%%%%%%%%%%%%%%%%%%%%%%%%%%%%%%%%%%%%%%%%%%%%%%%%%%%%%%%%%%%%%%%%%%%%%
\subsubsection{A simple $5d$ black hole with non-Abelian hair}
\label{sec-simple}
%%%%%%%%%%%%%%%%%%%%%%%%%%%%%%%%%%%%%%%%%%%%%%%%%%%%%%%%%%%%%%%%%%%%%%
%%%%%%%%%%%%%%%%%%%%%%%%%%%%%%%%%%%%%%%%%%%%%%%%%%%%%%%%%%%%%%%%%%%%%%
%%%%%%%%%%%%%%%%%%%%%%%%%%%%%%%%%%%%%%%%%%%%%%%%%%%%%%%%%%%%%%%%%%%%%%
%%%%%%%%%%%%%%%%%%%%%%%%%%%%%%%%%%%%%%%%%%%%%%%%%%%%%%%%%%%%%%%%%%%%%%

In order to add non-Abelian fields to our solutions it is exceedingly useful
to consider metrics with one additional isometry, because, then, we can make
use of our knowledge of the spherically symmetric solutions of the Bogomol'nyi
equations of the SU$(2)$ YMH system found by Protogenov in
Ref.~\cite{Protogenov:1977tq}. However, this isometry cannot be translational
if we want to find spherically-symmetric black holes because, then, the full
5-dimensional solution will have a translational isometry. Thus, we will start
with the choice $H=1/r$ ($r^{2}=y^{r}y^{r}$)\footnote{We need to distinguish
  between the Cartesian coordinates in $\mathbb{E}^{3}$, which we will denote
  by $y^{r}$ and the Cartesian coordinates in $\mathbb{E}^{4}$, which we will
  denote by $x^{m}$. The former are not a simple subset of the latter.} which,
as we have shown in Ref.~\cite{Bueno:2015wva}, relates the \textit{colored
  monopole} solution\footnote{This monopole is characterized by a vanishing
  magnetic charge.} to the the BPST instanton, which is spherically symmetric
in $\mathbb{E}^{4}$.

We are, thus, going to consider a configuration with the following
non-vanishing functions:

\begin{equation}
H = \frac{1}{r}\, , 
\hspace{0.3cm} 
L_{0} = A_{0}+\frac{q_{0}}{4r}\, , 
\hspace{0.3cm} 
L_{1} = A_{1}+\frac{q_{1}}{4r} \, , 
\hspace{0.3cm} 
\Phi^{A} = -f(r)\delta^{A}{}_{r} y^{r}\, ,
\end{equation}

\noindent
where $q_{0},q_{1}$ are electric charges in some convenient normalization,
$A_{0},A_{1}$ are constants to be determined through the normalization of the
metric and the scalar fields at infinity and $f(r)$ is the function (not to be
mistaken by $\hat{f}$) that characterizes the Higgs field in the
spherically-symmetric monopole solutions of
Ref.~\cite{Protogenov:1977tq}\footnote{See Appendix~\ref{sec-Protogenov} in
  which we have written all of Protogenov's solutions.}).

The next step consists in finding the 1-forms $\chi,\breve{A}^{I},\omega$ and
functions $L_{I}$ that satisfy
Eqs.~(\ref{eq:Hequation})-(\ref{eq:omegaequation2}) for the above
non-vanishing functions. $\omega$ is closed and can be set to zero, the
functions $L_{I}$ can also be set to zero while\footnote{The choice of angular
  coordinates is conditioned by the relation between the monopole and
  instanton as explained in Ref.~\cite{Bueno:2015wva}. We will identify the
  compact coordinate $z$ with the angular coordinate $\varphi$.}

\begin{equation}
\chi=d\varphi+\cos{\theta}d\psi\, ,
\hspace{1cm}
\breve{A}^{A}
=  
h(r)\varepsilon^{A}{}_{\underline{r}\underline{s}}y^{r}dy^{s}\, ,
\end{equation}

\noindent
where $h(r)$ is the function that characterizes the gauge field of the
monopole solution (see Appendix~\ref{sec-Protogenov})). The spacetime metric
is, then,

\begin{equation}
ds^{2} = \hat{f}^{\, 2}dt^{2} - \hat{f}^{\, -1}\left[ r(d\varphi+\cos{\theta}d \psi)^{2}
  +\frac{1}{r}(dr^{2} +r^{2}d\Omega^{2}_{(2)})\right]\, ,
\end{equation}

\noindent
where

\begin{equation}
d\Omega^{2}_{(2)} = d\theta^{2} +\sin^{2}{\theta}d\psi^{2}\, ,      
\end{equation}

\noindent
and, upon the change of coordinates $r= \rho^{2}/4$, it becomes

\begin{equation}
ds^{2} 
= 
\hat{f}^{\, 2}dt^{2} - \hat{f}^{\, -1}dx^{m}dx^{m}\, ,
\,\,\,\,\,
\mbox{where}
\,\,\,\,\,
dx^{m}dx^{m} = d\rho^{2} +\rho^{2} d\Omega^{2}_{(3)}\, .    
\end{equation}

For this configuration, the metric function
Eq.~(\ref{eq:d5metricfunctionST24}) is given by

\begin{equation}
\label{eq:metricfunctionstaticinstantonsolution}
\hat{f}^{\, -1}
=
3\sqrt[3]{
\tfrac{1}{2}\left(L_{0} -\tfrac{4}{3}r^{3}f^{2} \right)
(L_{1})^{2}
}\, ,  
\end{equation}

\noindent
and it immediately follows that in order for the solution to be asymptotically
regular, the monopole must be the colored one for which
$r^{3}f_{\lambda}^{2}\sim 1/r$, because for all the rest $r^{3}f^{2}\sim
r$ (see Appendix~\ref{sec-Protogenov}).  With this choice,\footnote{We are
  going to study the consequences of the other choices in
  Section~\ref{sec-godel}.} as shown in
Ref.~\cite{Bueno:2015wva}\footnote{More specifically, the gauge field one gets
  is $\hat{A}^{A(+)}_{L}$.}, the gauge field
$\hat{A}^{A}=\hat{A}^{A}{}_{\underline{m}}dx^{m}$ that follows from the use of
Eq.~(\ref{eq:Kromheimer}) is that of a BPST instanton in $\mathbb{E}^{4}$:

\begin{equation}
\hat{A}^{A}
 =  
\tfrac{1}{\tilde{g}}
\frac{1}{1+\lambda^{2}\rho^{2}/4}\,  v_{L}^{A}\, ,   
\end{equation}

\noindent
where $v_{L}^{A}$ are the SU$(2)$ left-invariant Maurer-Cartan
1-forms\footnote{In our conventions, these are given by 
\begin{equation}
\label{eq:MCcomponents}
\left\{
 \begin{array}{rcl}
v_{L}^{1} 
& = & 
\sin\psi\, d\theta -\sin\theta \cos\psi\, d\varphi\, ,
\\[2pt]
v_{L}^{2} 
& = & 
-\cos\psi\, d\theta -\sin\theta \sin\psi\, d\varphi\, ,
\\[2pt]
v_{L}^{3} 
& = & 
-(d\psi +\cos\theta\, d\varphi)\, ,
\\
 \end{array}
\right.
\hspace{1cm}
\mbox{and}
\hspace{1cm}
dv_{L}^{A}
+
\tfrac{1}{2}\epsilon_{ABC}\,
v_{L}^{B}
\wedge 
v_{L}^{C}
=
0\, .
\end{equation}
}. Since
the scalar functions $h^{A}$ vanish for this configuration, the full
5-dimensional vector fields are, according to
Eq.~(\ref{eq:completevectorfields}), given by

\begin{equation}
\begin{array}{rcl}
A^{0}
& = &
\frac{3^{5/2}}{2} (L_{1})^{2} \hat{f}^{\, 3} dt\, ,
\\
& & \\
A^{1}
& = &
3^{5/2} L_{1}\left(L_{0} -\tfrac{4}{3}r^{3}f_{\lambda}^{2} \right)\hat{f}^{\, 3} dt\, ,
\\
& & \\
A^{A}
& = &
{\displaystyle
\tfrac{1}{\tilde{g}}\frac{1}{1+\lambda^{2}\rho^{2}/4}
}\,  v_{L}^{A}\, .   
\\
\end{array}
\end{equation}

Finally, the only non-vanishing scalar is given by by

\begin{equation}
\phi \equiv h_{1}/h_{0}= \frac{L_{1}}{L_{0} -\tfrac{4}{3}r^{3}f_{\lambda}^{2}}\, .
\end{equation}

The integration constants are readily identified in terms of the asymptotic
value of the scalar as

\begin{equation}
A_{0} = \tfrac{2^{1/3}}{3} \phi_{\infty}^{-2/3}\, ,
\hspace{1cm}  
A_{1} = \tfrac{2^{1/3}}{3} \phi_{\infty}^{1/3}\, ,
\end{equation}

\noindent
while the mass and the area of the event horizon are given by 

\begin{eqnarray}
M
& = &  
2^{-1/3}3^{1/2} 
\left[
\phi_{\infty}^{2/3}q_{0} +2 \phi_{\infty}^{-1/3}q_{1}
\right]\, ,
\\
& & \nonumber \\
\frac{A}{2\pi^{2}}
& = &
\sqrt{\tfrac{3^{3}}{2}\left(q_{0} -\frac{2^{7}}{\tilde{g}^{2}}\right)(q_{1})^{2} }\, .
\end{eqnarray}

This solution can be understood as the result of the addition of a BPST
instanton to a standard 2-charge Abelian solution. This addition does not
produce any observable effects at spatial infinity, like, for instance, a
change in the mass, but does produce a change in the near-horizon geometry and
in the entropy.

The metric function of the 4-dimensional solution $e^{-2U}$ that one obtains
by dimensional reduction is related to the metric function of the
5-dimensional solution by

\begin{equation}
e^{-4U} = \frac{1}{r}\hat{f}^{-3}\, , 
\end{equation}

\noindent
which implies that the 4- and 5-dimensional solutions cannot be asymptotically
flat at the same time. In particular, with the choice made above
(corresponding to a colored monopole in $d=4$) $e^{-2u}\sim r^{-1/2}$ at
spatial infinity, a behavior that does not correspond to any known
vacuum. With the monopoles we discarded, however, we get an
asymptotically-flat solution. The near-horizon behavior is simultaneously good
in $d=4$ and $d=5$.

%%%%%%%%%%%%%%%%%%%%%%%%%%%%%%%%%%%%%%%%%%%%%%%%%%%%%%%%%%%%%%%%%%%%%%
%%%%%%%%%%%%%%%%%%%%%%%%%%%%%%%%%%%%%%%%%%%%%%%%%%%%%%%%%%%%%%%%%%%%%%
%%%%%%%%%%%%%%%%%%%%%%%%%%%%%%%%%%%%%%%%%%%%%%%%%%%%%%%%%%%%%%%%%%%%%%
%%%%%%%%%%%%%%%%%%%%%%%%%%%%%%%%%%%%%%%%%%%%%%%%%%%%%%%%%%%%%%%%%%%%%%
\subsubsection{A rotating $5d$ black hole with non-Abelian hair}
\label{sec-rotating}
%%%%%%%%%%%%%%%%%%%%%%%%%%%%%%%%%%%%%%%%%%%%%%%%%%%%%%%%%%%%%%%%%%%%%%
%%%%%%%%%%%%%%%%%%%%%%%%%%%%%%%%%%%%%%%%%%%%%%%%%%%%%%%%%%%%%%%%%%%%%%
%%%%%%%%%%%%%%%%%%%%%%%%%%%%%%%%%%%%%%%%%%%%%%%%%%%%%%%%%%%%%%%%%%%%%%
%%%%%%%%%%%%%%%%%%%%%%%%%%%%%%%%%%%%%%%%%%%%%%%%%%%%%%%%%%%%%%%%%%%%%%

In the context of timelike supersymmetric solutions of $\mathcal{N}=1,d=5$
supergravity rotation can be added by switching on the harmonic function $M$
\cite{Herdeiro:2002ft}. More specifically, we add to the static solution we
just constructed the harmonic function

\begin{equation}
M = \frac{J/2}{4r}\, , 
\end{equation}

\noindent
which only appears in Eq.~(\ref{eq:omega5}). The metric of the new solution
is

\begin{equation}
ds^{2} 
= 
\hat{f}^{\, 2}
\left[dt+\frac{J/2}{4r}(d\varphi+\cos{\theta}d \psi)\right]^{2} 
- \hat{f}^{\, -1}\left[ r(d\varphi+\cos{\theta}d \psi)^{2}
  +\frac{1}{r}(dr^{2} +r^{2}d\Omega^{2}_{(2)})\right]\, ,
\end{equation}

\noindent
where the metric function $\hat{f}$ is still given by
Eq.~(\ref{eq:metricfunctionstaticinstantonsolution}). The scalar field $\phi$
and the non-Abelian vector field $A^{A}$ take the same value as in the static
solution while the two Abelian vector fields are modified by the change 

\begin{equation}
dt \longrightarrow   dt+\frac{J/2}{4r}(d\varphi+\cos{\theta}d \psi)\, ,
\end{equation}

\noindent
which describes the presence of a magnetic dipole moment associated to the
rotation.

Asymptotically, the only novelty is the off-diagonal term $\sim
J/\rho^{2}dt(d\varphi+\cos{\theta}d \psi)$ which corresponds to identical values of the two
Casimirs of the angular momentum, both proportional to $J$, so this solution is a non-Abelian generalization of the Breckenridge--Myers-Peet--Vafa (BMPV) spinning black hole
 \cite{Breckenridge:1996is,Gauntlett:1998fz}. The mass has the
same expression in terms of the charges as in the static case.

In the near-horizon limit, if the behavior of the metric function $\hat{f}$ is

\begin{equation}
\hat{f}^{-1}\sim R^{2}/r\, ,  
\end{equation}

\noindent
for some constant $R$, the metric can be rewritten in the form

\begin{equation}
ds^{2}  
\sim
R^{2}d\Pi^{2}_{(2)} -R^{2}d\Omega^{2}_{(2)}  
-R^{2} 
\left[\cos{\alpha}(d\varphi +\cos{\theta}d\psi) -\sin{\alpha} \frac{r}{R^{2}}d\phi 
\right]^{2}\, ,
\end{equation}

\noindent
where $\phi$ is the rescaled time coordinate, defined as follows

\begin{equation}
\phi \equiv t/X\, ,
\hspace{1cm}
X/R \equiv \sqrt{1 - [J/(2R)^{3}]^{2}} \equiv \cos{\alpha}\, ,
\hspace{1cm}
(2R)^{3} \equiv 
\sqrt{\tfrac{3^{3}}{2}\left(q_{0} -\frac{2^{7}}{\tilde{g}^{2}}\right)(q_{1})^{2}
}\, , 
\end{equation}

\noindent
and $d\Pi^{2}_{(2)},d\Omega^{2}_{(2)}$ are the metrics of the 2-dimensional
Anti-de Sitter and sphere of unit radius

\begin{equation}
d\Pi^{2}_{(2)} \equiv  \left(\frac{r}{R^{2}}\right)^{2}d\phi^{2}
-\frac{dr^{2}}{r^{2}}\, .   
\end{equation}

The constant-time sections of the event horizon are squashed 3-spheres with
metric

\begin{equation}
-ds^{2}  
=
R^{2}
\left\{
\cos^{2}{\alpha}(d\varphi +\cos{\theta}d\psi)^{2}+d\Omega^{2}_{(2)}  
\right\}\, ,
\end{equation}

\noindent
and area

\begin{equation}
\frac{A}{2\pi^{2}}
 = 
\sqrt{\tfrac{3^{3}}{2}\left(q_{0} -\frac{2^{7}}{\tilde{g}^{2}}\right)(q_{1})^{2} -J^{2}}\, .
\end{equation}

%%%%%%%%%%%%%%%%%%%%%%%%%%%%%%%%%%%%%%%%%%%%%%%%%%%%%%%%%%%%%%%%%%%%%%
%%%%%%%%%%%%%%%%%%%%%%%%%%%%%%%%%%%%%%%%%%%%%%%%%%%%%%%%%%%%%%%%%%%%%%
%%%%%%%%%%%%%%%%%%%%%%%%%%%%%%%%%%%%%%%%%%%%%%%%%%%%%%%%%%%%%%%%%%%%%%
%%%%%%%%%%%%%%%%%%%%%%%%%%%%%%%%%%%%%%%%%%%%%%%%%%%%%%%%%%%%%%%%%%%%%%
\subsubsection{A more general solution}
\label{sec-godel}
%%%%%%%%%%%%%%%%%%%%%%%%%%%%%%%%%%%%%%%%%%%%%%%%%%%%%%%%%%%%%%%%%%%%%%
%%%%%%%%%%%%%%%%%%%%%%%%%%%%%%%%%%%%%%%%%%%%%%%%%%%%%%%%%%%%%%%%%%%%%%
%%%%%%%%%%%%%%%%%%%%%%%%%%%%%%%%%%%%%%%%%%%%%%%%%%%%%%%%%%%%%%%%%%%%%%
%%%%%%%%%%%%%%%%%%%%%%%%%%%%%%%%%%%%%%%%%%%%%%%%%%%%%%%%%%%%%%%%%%%%%%

In Section~\ref{sec-simple} we used the colored monopole solution in order to
obtain an asymptotically flat black-hole solution in the simplest
way. However, we can also use the monopoles in the 2-parameter family, for
which, asymptotically, $r^{3}f^{2}\sim r$ if we switch on additional harmonic
functions and choose the values of the integration constants appropriately so
that the metric functions $\hat{f}(r),\omega_{5},\omega$ give an
asymptotically-flat solution.

Throughout the following discussion, it is convenient to have the explicit
form of these functions for $H=1/r$, $\Phi^{A}=-f(r)\delta^{A}{}_{r}y^{r}$
and $L_{A}=0$ at hand:

\begin{equation}
  \begin{array}{rcl}
\hat{f}^{\, -3}
& = &
27
\left[
\tfrac{1}{2}L_{0} +\tfrac{2}{3}r[(\Phi^{1})^{2}-r^{2}f^{2}]
\right]
\left[
(L_{1})^{2} +\tfrac{16}{3} r\Phi^{0}L_{1}\Phi^{1}
+\tfrac{64}{9}(r\Phi^{0})^{2}[(\Phi^{1})^{2}-r^{2}f^{2}]
\right]\, ,
\\
& & \\
\omega_{5}
& = &
M+8\sqrt{2}\, r^{2} \Phi^{0}[(\Phi^{1})^{2}-r^{2}f^{2}] 
+3\sqrt{2}\, r L_{i} \Phi^{i} \, ,
\\
& & \\
\star_{3} d \omega
& = &
\frac{1}{r} dM-Md\frac{1}{r}
+3\sqrt{2} \left( \Phi^{i} dL_{i}-L_{i}d\Phi^{i} \right)\, ,
\end{array}
\end{equation}

\noindent
where $i=0,1$. Apart from the functions $H$ and $\Phi^{A}$, we are going to
consider the following non-vanishing harmonic functions

\begin{equation}
  \{\Phi^{0},\Phi^{1},L_{0},L_{1},M\}\, ,
\end{equation}

\noindent
with

\begin{equation}
\Phi^{0,1} = A^{0,1} +\frac{p^{0,1}}{4r}\, ,
\hspace{1cm}  
L_{0,1} = A_{0,1} +\frac{q_{0,1}}{4r}\, ,
\hspace{1cm}
M= a+\frac{b}{4r}\, .  
\end{equation}

$\hat{f}^{-3}$ is a product of two factors. Our strategy will be to make the
constant piece of $\Phi^{1}$, $A^{1}$, cancel the constant piece in $rf(r)$,
$\mu/g$ so that $[(\Phi^{1})^{2}-r^{2}f^{2}]$ is asymptotically
$\mathcal{O}(1/r)$\footnote{We choose the positive sign for simplicity.}:

\begin{equation}
A^{1}= \mu/g\,.  
\end{equation}

\noindent
This ensures that the second term in $\hat{f}^{-3}$ diverges asymptotically at
most as $\mathcal{O}(r)$ while the first is asymptotically constant. This
constant can be made to vanish by choosing the constant piece of $L_{0}$,
$A_{0}$, to be

\begin{equation}
A_{0} 
=      
-\tfrac{8}{3} 
\frac{\mu}{g}\left(\frac{1}{g}+\frac{p^{1}}{4} \right)
\, ,  
\end{equation}

\noindent
and now the first term is asymptotically $\mathcal{O}(1/r)$ and $\hat{f}^{-3}$
is asymptotically constant.

Next, we require that all the $\mathcal{O}(r^{2})$, $\mathcal{O}(r)$ and
$\mathcal{O}(1)$ terms in $\omega_{5}$ vanish\footnote{Observe that this does
  not imply the complete vanishing of $\omega_{5}$: there are
  $\mathcal{O}(1/r)$ terms that give angular momentum (which could be
  cancelled by the integration constant $b$ in $M$) and also
  $\mathcal{O}(e^{-4\mu r})$ terms that cannot be cancelled. Therefore, the
  metric is not static even if the angular momentum is set to zero.}. This
gives two new relations\footnote{The above values of $A_{0}$ and $A^{1}$ make
  the $\mathcal{O}(r^{2})$ term vanish.} between the constants $A_{i}$,
$A^{i}$ and $a$. The vanishing of $\omega$ gives another relation between the
same constants. Thus, requiring asymptotic flatness fixes the values of all
these constants in terms of the Abelian charges $p^{i},q_{i}$ and $\mu$ and
$g$. Finally the normalization of the metric at infinity also fixes the value
of $\mu$ and the solution has no free moduli!

The values of the integration constants $A_{0},A^{1}$ has been given above and
the values of the rest are\footnote{We have not reexpressed the 4-dimensional
  gauge coupling constant $g$ in terms of the 5-dimensional, $\tilde{g}$ to
  have simpler expressions.}

\begin{equation}
\begin{array}{rcl}
A_{1}
& = &
-\frac{88}{3}A^{0} 
{\displaystyle
\left(\frac{1}{g}+ \frac{p^{1}}{4}\right)
}\, ,
\\
& & \\
A^{0}
& = &
{\displaystyle
\left\{
\frac
{
\left(16p^{0}+4gp^{0}p^{1}+gq^{1}\right)
\left(4+gp^{1}\right)^{-1}
}
{
40
\left(3q_{0}+(p^{1})^{2}-\frac{16}{g^{2}} \right)
\left(q_{0}+2(p^{1})^{2}-\frac{32}{g^{2}} \right)
}
\right\}^{1/3}
}\, ,
\\
& & \\
\mu
& = & 
A^{0}
\left[
{\displaystyle
\frac{32-2g^{2}(p^{1})^{2}-g^{2}q_{0}}
{16p^{0}+4g p^{0} p^{1}+g q_{1}}
}
\right]\, , 
\\
& & \\
a
& = & 
{\displaystyle
\sqrt{2}A^{0}
\left[ 
\frac{48}{g^{2}}
+\frac{22p^{1}}{g}
+\frac{5(p^{1})^{2}}{2}
-\frac{3q_{0}}{4}
\right]
-\sqrt{2}
\left[ 
\frac{22\mu p^{0}}{g^{2}}
+\frac{11 \mu p^{0} p^{1}}{2g}
+ \frac{3\mu q_{1}}{4g}
\right]
}\, ,
\\
& & \\
{\displaystyle
b
}
& = &
{\displaystyle
J/2
-
6\sqrt{2}
\left[
\frac{p^{0}(p^{1})^{2}}{2}
+\frac{p^{0}q_{0}+p^{1}q^{1}}{8}
-8\frac{p^{0}}{g^{2}}
\right]
}\, ,
\\
\end{array}
\end{equation}

\noindent
where $J$ is the angular momentum.

The mass of this solution is given by

\begin{equation}
M 
= 
\frac{\pi A^{0}}{2 G}
\left[
3q_{0}+(p^{1})^{2}-\frac{16}{g^{2}}
\right]
\left[
3\frac{\mu}{g}q_{1}
+8\left(\frac{1}{g}+\frac{p^{1}}{4} \right)
\left(10A^{0} \left( \frac{24}{g}+5p^{1} \right)-9\frac{\mu}{g}p^{0} \right)
\right]
\, .
\end{equation}

\noindent
and the area of the horizon is

\begin{equation}
\frac{A}{2\pi^{2}}
=
\sqrt{
\tfrac{1}{2} 
\left[
3q_{0}+(p^{1})^{2}-\frac{16}{g^{2}}
\right] 
\left[
3q_{1}+2p^{1}p^{0}-\frac{8p^{0}}{g}
\right]
\left[
3q_{1}+2p^{0}p^{1}+\frac{8p^{0}}{g}
\right] 
-J^{2}}\, .
\end{equation}

%%%%%%%%%%%%%%%%%%%%%%%%%%%%%%%%%%%%%%%%%%%%%%%%%%%%%%%%%%%%%%%%%%%%%%
%%%%%%%%%%%%%%%%%%%%%%%%%%%%%%%%%%%%%%%%%%%%%%%%%%%%%%%%%%%%%%%%%%%%%%
%%%%%%%%%%%%%%%%%%%%%%%%%%%%%%%%%%%%%%%%%%%%%%%%%%%%%%%%%%%%%%%%%%%%%%
%%%%%%%%%%%%%%%%%%%%%%%%%%%%%%%%%%%%%%%%%%%%%%%%%%%%%%%%%%%%%%%%%%%%%%
\subsubsection{Null supersymmetric non-Abelian $5d$ solutions from $4d$ black
  holes and global monopoles}
\label{sec-nullfrombhs}

%%%%%%%%%%%%%%%%%%%%%%%%%%%%%%%%%%%%%%%%%%%%%%%%%%%%%%%%%%%%%%%%%%%%%%
%%%%%%%%%%%%%%%%%%%%%%%%%%%%%%%%%%%%%%%%%%%%%%%%%%%%%%%%%%%%%%%%%%%%%%
%%%%%%%%%%%%%%%%%%%%%%%%%%%%%%%%%%%%%%%%%%%%%%%%%%%%%%%%%%%%%%%%%%%%%%
%%%%%%%%%%%%%%%%%%%%%%%%%%%%%%%%%%%%%%%%%%%%%%%%%%%%%%%%%%%%%%%%%%%%%%

Using the general results of the preceding sections it is very easy to
construct null supersymmetric solutions by uplifting 4-dimensional timelike
supersymmetric solutions with $\mathcal{I}^{0}$. In particular, we can uplift
the black-hole and global-monopole solutions of the ST$[2,5]$ model recently
constructed in Ref.~\cite{Bueno:2014mea}. In this paper we will focus on the
single center solutions only. 

The 4-dimensional solutions depend on the following non-vanishing
$\mathcal{I}^{M}$

\begin{equation}
\begin{array}{rclrclrcl}
\mathcal{I}^{1} 
& = & 
A^{1}+{\displaystyle\frac{p^{1}/\sqrt{2}}{r}}\, , \hspace{.7cm}&
\mathcal{I}^{2} 
& = & 
A^{2}+{\displaystyle\frac{p^{2}/\sqrt{2}}{r}}\,  ,\hspace{.7cm}&
\mathcal{I}^{A} 
& = & 
\sqrt{2}\, \delta^{A}{}_{p}x^{p}f(r)\, ,\\
& & & & & & & & \\
\mathcal{I}_{0} 
& = & 
A_{0}+{\displaystyle\frac{q_{0}/\sqrt{2}}{r}}\, ,
& & & & & & \\
\end{array}
\end{equation}

\noindent
where $f(r)$ is the function $f_{\mu,s}$ or $f_{\lambda}$ in
Appendix~\ref{sec-Protogenov} corresponding to one of the
spherically-symmetric BPS SU$(2)$ monopoles, $p^{1},p^{2},q_{0}$ are magnetic
and electric charges and $A^{1},A^{2},A_{0}$ integration constants to be
determined in terms of the asymptotic values of the scalars and the metric.

The 5-dimensional metric is that of an intersection of a string lying along
the $z$ direction and a $pp$-wave propagating along the same direction:

\begin{equation}
ds^{2} = 2\ell du(dv +Kdu) - \ell^{-2}d\vec{x}^{2}_{(3)}\, ,  
\end{equation}

\noindent
where

\begin{equation}
\ell^{-3} = 4 \mathcal{I}^{1} [(\mathcal{I}^{2})^{2} -2r^{2}f^{2}]\, ,
\hspace{1cm}
K= 1+2\mathcal{I}_{0}\, .  
\end{equation}

The scalar fields, defined by $\phi^{x} \equiv h^{x}/h^{0}$, are given by 

\begin{equation}
\phi^{1} = \mathcal{I}^{2}/\mathcal{I}^{1}\, ,
\hspace{1cm}
\phi^{A} = -\delta^{A}{}_{p}x^{p}f(r)/\mathcal{I}^{1}\, ,  
\end{equation}

\noindent
and the vector fields are given by

\begin{equation}
A^{0,1} = -2\sqrt{6} p^{1,2} A\, ,
\hspace{1cm}  
A^{A} = 2\sqrt{6} h(r)\epsilon^{A}{}_{rs}x^{r}dx^{s}\, ,
\end{equation}

\noindent
where $A$ is the vector field of a Dirac magnetic monopole of unit charge,
satisfying $dA=\star_{3}d \frac{1}{r}$ and $h(r)$ is the function $h_{\mu,s}$
or $h_{\lambda}$ in Appendix~\ref{sec-Protogenov} corresponding to one of the
spherically-symmetric BPS SU$(2)$ monopoles.

The 4-dimensional electric charge $q_{0}$ corresponds to the momentum of the
5-dimensional gravitational wave in the $z$ direction and none of the scalar
and vector fields depend on it. For the sake of simplicity we are going to set
it to zero ($q_{0}=0$ and $\mathcal{I}_{0}=-1/2$ so $K=0$) and we are going to
analyze the string solutions with the above scalar and vector fields and with
metric

\begin{equation}
ds^{2} = \ell (dt^{2}-dz^{2}) - \ell^{-2}d\vec{x}^{2}_{(3)}\, ,  
\end{equation}

\noindent
with the metric function $\ell$ given as above.

The metric will be regular in the $r\rightarrow 0 $ limit if $\ell \sim r $ or
$\ell \sim \mathrm{constant}$. These two behaviors are, respectively, those of
extremal black strings in the near-horizon limit and those of global
monopoles. Let us consider each case separately.

\begin{description}
\item[Global string-monopoles] These are the string-like solutions that, upon
  dimensional reduction along $z$, give the spherically-symmetric global
  monopoles constructed in Ref.~\cite{Bueno:2014mea}. They can be constructed
  with $f(r)= f_{\mu,s=0}(r)$ (the BPST 't Hooft-Polyakov monopole) and with
  $p^{1}=p^{2}=0$, so that

\begin{equation}
\ell^{-3} = 4 A^{1} [(A^{2})^{2} -2r^{2}f_{\mu,\,  s=0}^{2}]\, ,
\hspace{.5cm}
\phi^{1} = A^{2}/A^{1}\, ,
\hspace{.5cm}
\phi^{A} = -\sqrt{2} \delta^{A}{}_{r}x^{r} f_{\mu,s=0}(r)/A^{1}\, ,
\end{equation}

\noindent
and the only non-trivial vector field is $A^{A}$. 

The integration constants $A^{1,2},\mu$ are given by

\begin{equation}
A^{1} = \frac{1}{\chi_{\infty}^{1/3}}\, , 
\hspace{.5cm}
A^{2} = \frac{\phi^{1}_{\infty}}{\chi_{\infty}^{1/3}}\, ,
\hspace{.5cm}
\mu = \frac{g|\phi_{\infty}|}{\sqrt{2}\chi_{\infty}^{1/3}}\, ,
\hspace{.5cm}
\chi_{\infty} \equiv 4[(\phi^{1}_{\infty})^{2} -|\phi_{\infty}|^{2}]\, ,   
\end{equation}

\noindent
where $|\phi_{\infty}|^{2}$ is the asymptotic value of the gauge-invariant
combination $\phi^{A}\phi^{A}$, and the string's tension (simply defined as
minus the coefficient of $1/r$ in the large-$r$ expansion of $g_{tt}$) is
given by \cite{deAntonioMartin:2012bi}

\begin{equation}
T_{\rm monopole} = \frac{32 |\phi_{\infty}|}{\sqrt{3}\chi_{\infty}^{2/3}}
\frac{1}{|\tilde{g}|}\, .  
\end{equation}

These are globally regular solutions with no horizons, like their
4-dimensional analogues.

\item[Black strings] They must necessarily have non-vanishing magnetic charges
  $p^{1,2}$ in order to have a regular horizon. This horizon will be a
  2-dimensional surface characterized by being normal to 2 linearly
  independent null vectors. The mass and entropy of the black string will
  depend on the choice of monopole.

  Let us first consider the BPST 't Hooft-Polyakov monopole (or equivalently,
  let us add magnetic charges $p^{1,2}$ to the above global monopole). In this
  case, the relation between the integration constants $A^{1,2},\mu$ and the
  asymptotic values of the scalars will be the same as before. The string's
  tension and the area of the horizon contain contributions from the magnetic
  charges $p^{1},p^{2}$:

\begin{eqnarray}
T 
& = &
\tfrac{1}{3\sqrt{2}}\chi_{\infty}^{1/3} 
\left[p^{1} +8 \frac{\phi^{1}_{\infty}}{\chi_{\infty}}p^{2} \right] 
+T_{\rm monopole}\, ,
\\
& & \nonumber \\
\frac{A}{4\pi}
& = &
2\left[ p^{1}(p^{2})^{2}\right]^{2/3}\, .
\end{eqnarray}

When we consider the more general 't Hooft-Polyakov-Protogenov monopole we find that the area of the horizon receives a contribution from the non-Abelian charge,
\begin{equation}
\frac{A}{4\pi}
= 
2\left\{ p^{1}\left[(p^{2})^{2}-\frac{2}{g^{2}}\right]\right\}^{2/3}\, .
\end{equation}

\end{description}

%%%%%%%%%%%%%%%%%%%%%%%%%%%%%%%%%%%%%%%%%%%%%%%%%%%%%%%%%%%%%%%%%%%%%%
%%%%%%%%%%%%%%%%%%%%%%%%%%%%%%%%%%%%%%%%%%%%%%%%%%%%%%%%%%%%%%%%%%%%%%
%%%%%%%%%%%%%%%%%%%%%%%%%%%%%%%%%%%%%%%%%%%%%%%%%%%%%%%%%%%%%%%%%%%%%%
%%%%%%%%%%%%%%%%%%%%%%%%%%%%%%%%%%%%%%%%%%%%%%%%%%%%%%%%%%%%%%%%%%%%%%
\section{Conclusions}
\label{sec-conclusions}
%%%%%%%%%%%%%%%%%%%%%%%%%%%%%%%%%%%%%%%%%%%%%%%%%%%%%%%%%%%%%%%%%%%%%%
%%%%%%%%%%%%%%%%%%%%%%%%%%%%%%%%%%%%%%%%%%%%%%%%%%%%%%%%%%%%%%%%%%%%%%
%%%%%%%%%%%%%%%%%%%%%%%%%%%%%%%%%%%%%%%%%%%%%%%%%%%%%%%%%%%%%%%%%%%%%%
%%%%%%%%%%%%%%%%%%%%%%%%%%%%%%%%%%%%%%%%%%%%%%%%%%%%%%%%%%%%%%%%%%%%%%

In this paper we have studied the general procedure to construct timelike and
null supersymmetric solutions of $\mathcal{N}=1,d=5$ SEYM theories that can be
dimensionally reduced to timelike solutions of $\mathcal{N}=2,d=4$ SEYM
theories. These solutions, therefore, can also be constructed by oxidation of
the 4-dimensional solutions and we have striven to clarify this procedure and
find the relations between the 4- and 5-dimensional fields and the 4- and
5-dimensional equations they satisfy. The relation between instantons in
4-dimensional hyperK\"ahler spaces and monopoles satisfying the Bogomol'nyi
equation in $\mathbb{E}^{3}$ found by Kronheimer plays a crucial role in this
relation and, in combination with the results obtained in
Ref.~\cite{Bueno:2015wva}, it allows us to construct spherically-symmetric
5-dimensional solutions that contain YM instantons. The standard oxidation of
monopoles gives rise to 5-dimensional solutions that have an additional
translational isometry and cannot be spherically symmetric.

We have exploited the general results to construct the first 5-dimensional
black-hole and black-string solutions with non-Abelian YM fields. The simplest
black-hole solutions contain the field of a BPST instanton in the so-called
\textit{base space} and their behavior is similar to that of the colored black
holes found in 4-dimensional SEYM theories
\cite{Meessen:2008kb,Meessen:2015nla}: the non-Abelian YM field cannot be
``seen'' at spatial infinity, it does not contribute to the mass, but it can
be seen in the near-horizon limit and it contributes to the entropy. One can
compare the entropies of the simplest non-Abelian black hole with that of
another black hole with the same Abelian charges and moduli (and, henceforth,
with the same mass). The entropy of the former is always smaller, so it is
entropically favorable to lose the non-Abelian field. It is not clear by which
mechanism this can happen.

We have also found more complicated black-hole solutions which contain the
field of the instantons that one obtains by reducing Protogenov monopoles in
the so-called \textit{base space}. Those instantons are not regular in flat
space and, in general, the spacetime metrics they give rise to are not
asymptotically flat. We have shown that a judicious choice of the integration
constants (and, hence, of the moduli) in terms of the charges produces a
metric that is not only asymptotically flat with positive mass but also has a
regular horizon. Thus, at special points in the moduli space of the scalar
manifold, additional non-Abelian black-hole solutions are possible. In these
solutions, the YM fields do contribute to the mass and to the entropy.

Finally, we have also found black-string solutions by conventional oxidation
of non-Abelian black-hole solutions from 4 dimensions. One of them is a
globally-regular string-monopole solution and the rest are more conventional
solutions.

It is clear that the new solutions that we have constructed need further
study. Their string-theoretic interpretation could be very interesting. The
model we have chosen to construct explicit solutions is a truncation of the
effective theory of the heterotic string compactified to 5 dimensions and can,
alternatively, be seen as associated to the compactification of the type~IIB
theory in K3 times a circle. This should simplify a bit the task and, perhaps,
open the way to a microscopic interpretation of entropies that depend on
parameters that do not appear at infinity. Work in this direction is in
progress.

%%%%%%%%%%%%%%%%%%%%%%%%%%%%%%%%%%%%%%%%%%%%%%%%%%%%%%%%%%%%%%%%%%%%%%
%%%%%%%%%%%%%%%%%%%%%%%%%%%%%%%%%%%%%%%%%%%%%%%%%%%%%%%%%%%%%%%%%%%%%%
%%%%%%%%%%%%%%%%%%%%%%%%%%%%%%%%%%%%%%%%%%%%%%%%%%%%%%%%%%%%%%%%%%%%%%
%%%%%%%%%%%%%%%%%%%%%%%%%%%%%%%%%%%%%%%%%%%%%%%%%%%%%%%%%%%%%%%%%%%%%%
\section*{Acknowledgments}
%%%%%%%%%%%%%%%%%%%%%%%%%%%%%%%%%%%%%%%%%%%%%%%%%%%%%%%%%%%%%%%%%%%%%%
%%%%%%%%%%%%%%%%%%%%%%%%%%%%%%%%%%%%%%%%%%%%%%%%%%%%%%%%%%%%%%%%%%%%%%
%%%%%%%%%%%%%%%%%%%%%%%%%%%%%%%%%%%%%%%%%%%%%%%%%%%%%%%%%%%%%%%%%%%%%%
%%%%%%%%%%%%%%%%%%%%%%%%%%%%%%%%%%%%%%%%%%%%%%%%%%%%%%%%%%%%%%%%%%%%%%

The authors would like to thank A.~Anabal\'on for interesting conversations.
This work has been supported in part by the Spanish Ministry of Science and
Education grants FPA2012-35043-C02 (-01 {\&} -02), the Centro de Excelencia
Severo Ochoa Program grant SEV-2012-0249, the EU-COST Action MP1210 ``The
String Theory Universe'', the Principado de Asturias grant GRUPIN14-108 and
the Spanish Consolider-Ingenio 2010 program CPAN CSD2007-00042.  The work was
further supported by the \textit{Severo Ochoa} pre-doctoral grant
SVP-2013-067903 (PF-R).  TO wishes to thank M.M.~Fern\'andez for her permanent
support.

%%%%%%%%%%%%%%%%%%%%%%%%%%%%%%%%%%%%%%%%%%%%%%%%%%%%%%%%%%%%%%%%%%%%%%
%%%%%%%%%%%%%%%%%%%%%%%%%%%%%%%%%%%%%%%%%%%%%%%%%%%%%%%%%%%%%%%%%%%%%%
%%%%%%%%%%%%%%%%%%%%%%%%%%%%%%%%%%%%%%%%%%%%%%%%%%%%%%%%%%%%%%%%%%%%%%
%%%%%%%%%%%%%%%%%%%%%%%%%%%%%%%%%%%%%%%%%%%%%%%%%%%%%%%%%%%%%%%%%%%%%%
\appendix
%%%%%%%%%%%%%%%%%%%%%%%%%%%%%%%%%%%%%%%%%%%%%%%%%%%%%%%%%%%%%%%%%%%%%%
%%%%%%%%%%%%%%%%%%%%%%%%%%%%%%%%%%%%%%%%%%%%%%%%%%%%%%%%%%%%%%%%%%%%%%
%%%%%%%%%%%%%%%%%%%%%%%%%%%%%%%%%%%%%%%%%%%%%%%%%%%%%%%%%%%%%%%%%%%%%%
%%%%%%%%%%%%%%%%%%%%%%%%%%%%%%%%%%%%%%%%%%%%%%%%%%%%%%%%%%%%%%%%%%%%%%
\section{Dimensional reduction of $\mathcal{N}=1,d=5$ SEYM theories}
\label{ap-reduction}
%%%%%%%%%%%%%%%%%%%%%%%%%%%%%%%%%%%%%%%%%%%%%%%%%%%%%%%%%%%%%%%%%%%%%%
%%%%%%%%%%%%%%%%%%%%%%%%%%%%%%%%%%%%%%%%%%%%%%%%%%%%%%%%%%%%%%%%%%%%%%
%%%%%%%%%%%%%%%%%%%%%%%%%%%%%%%%%%%%%%%%%%%%%%%%%%%%%%%%%%%%%%%%%%%%%%
%%%%%%%%%%%%%%%%%%%%%%%%%%%%%%%%%%%%%%%%%%%%%%%%%%%%%%%%%%%%%%%%%%%%%%

$\mathcal{N}=1,d=5$ supergravity coupled to vector multiplets gives
$\mathcal{N}=2,d=4$ supergravity coupled to vector multiplets upon dimensional
reduction over a spacelike circle\footnote{See, for instance,
  Refs.~\cite{Freedman:2012zz} and references therein}. If some non-Abelian
subgroup of the isometry group of the scalar manifold of the 5-dimensional
theory has been gauged, and we perform a simple (as opposed to a generalized)
dimensional reduction, the 4-dimensional theory will have exactly the same
non-Abelian subgroup of the (now bigger) isometry group gauged. Thus
$\mathcal{N}=1,d=5$ and $\mathcal{N}=2,d=4$ SEYM theories are related by
dimensional reduction over a spacelike circle.

It should be clear that, under the above conditions, the relation between the
5- and 4-dimensional fields in the gauged theories is exactly the same as in
the ungauged one and is, therefore, well known. In the conventions we follow
here\footnote{That is, the conventions used in
  Refs.~\cite{Bellorin:2006yr,Bellorin:2007yp,Bellorin:2008we} for the
  $\mathcal{N}=1,d=5$ theories and in the conventions used in
  Refs.~\cite{Meessen:2006tu,Huebscher:2006mr,Huebscher:2007hj,Meessen:2008kb,Hubscher:2008yz,Meessen:2012sr,Bueno:2014mea,Meessen:2015nla,Bueno:2015wva}
  for the $\mathcal{N}=2,d=4$ theories.} the relation between the bosonic
fields of an $\mathcal{N}=1,d=5$ supergravity model defined by $C_{IJK}$
(tilded) and the bosonic fields of a cubic model of $\mathcal{N}=2,d=4$
supergravity defined by the symmetric tensor $d_{ijk}$ (untilded) are
\footnote{See, for instance, Ref.~\cite{kn:GS} which follows the conventions
  used here.}

\begin{equation}
\label{eq:KKreductionN1d5N2d4}
\begin{array}{rclrcl}
g_{\mu\nu}
& = &
|\tilde{g}_{\underline{z}\underline{z}}|^{\frac{1}{2}}
\left(\tilde{g}_{\mu\nu}
-\tilde{g}_{\mu\underline{z}}
\tilde{g}_{\nu\underline{z}}/\tilde{g}_{\underline{z}\underline{z}}\right),\,\,\,\,\,
&
d_{ijk}
& = & 
6  C_{i-1\, j-1\, k-1},
\\
& & & & & \\
A^{0}{}_{\mu}
& = &
\tfrac{1}{2\sqrt{2}} 
\tilde{g}_{\mu\underline{z}}/\tilde{g}_{\underline{z}\underline{z}}\, ,
&
A^{i}{}_{\mu}  
& = &
-\tfrac{1}{2\sqrt{6}}
\left(
\tilde{A}^{i-1}{}_{\mu} -\tilde{A}^{i-1}{}_{\underline{z}}
\tilde{g}_{\mu\underline{z}}/\tilde{g}_{\underline{z}\underline{z}}
\right)\, ,
\\
& & & & & \\
Z^{i}
& = &
\tfrac{1}{\sqrt{3}}
\tilde{A}^{i-1}{}_{\underline{z}}
+i|\tilde{g}_{\underline{z}\underline{z}}|^{\frac{1}{2}}
\tilde{h}^{i-1}\, ,
&
& & \\
\end{array}
\end{equation}

\noindent
and the inverse relations are

\begin{equation}
\begin{array}{rclrcl}
\tilde{g}_{\underline{z}\underline{z}}
& = & 
-k^{2}\, ,
& 
\tilde{A}^{I}{}_{\underline{z}}
& = & 
\sqrt{3}\Re\mathfrak{e} Z^{I+1}\, ,
\\
& & & & & \\    
\tilde{g}_{\mu\underline{z}}
& = & 
-2\sqrt{2}k^{2}A^{0}{}_{\mu}\, ,
&
\tilde{A}^{I}{}_{\mu}
& = & 
-2\sqrt{6}\left(A^{I+1}{}_{\mu}-\Re\mathfrak{e} Z^{I+1}A^{0}{}_{\mu}\right)\, ,
\\
& & & & & \\
\tilde{g}_{\mu\nu}
& = & 
k^{-1}g_{\mu\nu} -8k^{2}A^{0}{}_{\mu}A^{0}{}_{\nu}\, ,
\hspace{1cm}
&
\tilde{h}^{I}
& = & 
k^{-1}\Im\mathfrak{m}Z^{I+1}\, .
\\
\end{array}
\end{equation}

In these relations it has been taken into account that, if $n_{v}$ denotes the
number of vector multiplets in $d=5$, then, the 4-dimensional theory has
$n_{v}+1$ vector multiplets so that $I,J,K=0,\cdots,n_{v}$,\,\, $i,j,k=0,\cdots,
n_{v}+1$. The additional 4-dimensional vector multiplet is the $i=0$ one and,
therefore, the 5-dimensional vector labeled by $I$ corresponds to the
4-dimensional vector labeled by $i=I+1$. 

While this is the whole story for the fields, it is important to realize that
the factor that related the 4- and 5-dimensional gauge fields changes the
standard form of the covariant derivatives and gauge field strengths and it
must be absorbed into a redefinition of the gauge coupling constant. Thus, we
also have

\begin{equation}
\tilde{g} = -2\sqrt{6} g\, .  
\end{equation}

Observe that this result has been obtained using the orientation
$\varepsilon^{0123z}=+1$, which is not the one we are using in the main text
($\varepsilon^{0z123}=+1$). However, in practice, the result can be adapted to
that orientation by reversing the sign of each $z$ tensor index. This
operation only changes the sign of $A^{0}{}_{\mu}$ and
$\Re\mathfrak{e} Z^{i}$.

%%%%%%%%%%%%%%%%%%%%%%%%%%%%%%%%%%%%%%%%%%%%%%%%%%%%%%%%%%%%%%%%%%%%%%
%%%%%%%%%%%%%%%%%%%%%%%%%%%%%%%%%%%%%%%%%%%%%%%%%%%%%%%%%%%%%%%%%%%%%%
%%%%%%%%%%%%%%%%%%%%%%%%%%%%%%%%%%%%%%%%%%%%%%%%%%%%%%%%%%%%%%%%%%%%%%
%%%%%%%%%%%%%%%%%%%%%%%%%%%%%%%%%%%%%%%%%%%%%%%%%%%%%%%%%%%%%%%%%%%%%%
\section{Spherically-symmetric  solutions of the 
$\mathrm{SU}(2)$ Bogomol'nyi equations in $\mathbb{E}^{3}$}
\label{sec-Protogenov}
%%%%%%%%%%%%%%%%%%%%%%%%%%%%%%%%%%%%%%%%%%%%%%%%%%%%%%%%%%%%%%%%%%%%%%
%%%%%%%%%%%%%%%%%%%%%%%%%%%%%%%%%%%%%%%%%%%%%%%%%%%%%%%%%%%%%%%%%%%%%%
%%%%%%%%%%%%%%%%%%%%%%%%%%%%%%%%%%%%%%%%%%%%%%%%%%%%%%%%%%%%%%%%%%%%%%
%%%%%%%%%%%%%%%%%%%%%%%%%%%%%%%%%%%%%%%%%%%%%%%%%%%%%%%%%%%%%%%%%%%%%%

The equations of motion of the SU$(2)$ Yang-Mills-Higgs (YMH) theory in the
Bogomol'nyi-Prasad-Sommerfield (BPS) limit in which the the Higgs potential
vanishes read

\begin{eqnarray}
\mathfrak{D}_{\mu} F^{A\, \mu\nu} & = &  
-g\varepsilon_{BC}{}^{A}\Phi^{B} \mathfrak{D}^{\nu}\Phi^{C}\, ,
\\
& & \nonumber \\
\mathfrak{D}^{2} \Phi^{A} & = & 0\, .  
\end{eqnarray}

Static configurations satisfying the first-order \textit{Bogomol'nyi
  equations} \cite{Bogomolny:1975de}

\begin{equation}
\label{eq:Beqs}
F^{A}{}_{\underline{r}\underline{s}} 
= 
\varepsilon_{rst}\mathfrak{D}_{\underline{t}}\Phi{A}\, ,
\end{equation}

\noindent
can be seen to satisfy all the above second-order YMH equations of motion. 

BPS magnetic monopole solutions such as the (BPS) 't Hooft-Polyakov monopole
found by Prasad and Sommerfield in Ref.~\cite{Prasad:1975kr} satisfy the
Bogomol'nyi equations and, therefore, it is of some interest to identify all
their solutions. In the spherically-symmetric case this problem was
solved by Protogenov in Ref.~\cite{Protogenov:1977tq} and his solution can be
described as follows: the Higgs and gauge field can always be brought to this
form (\textit{hedgehog ansatz})

\begin{equation}
\Phi^{A} = -\delta^{A}{}_{s}f(r)y^{s}\, ,
\hspace{1cm}
A^{A}{}_{\underline{r}} = -\varepsilon^{A}{}_{rs}y^{s} h(r)\,  ,
\end{equation}

\noindent
in which they are characterized by just two functions, $f(r),h(r)$ of the
radial coordinate $r=\sqrt{y^{s}y^{s}}$. There is only a 2-parameter family for
which these functions, denoted by $(f_{\mu,s},h_{\mu,s})$, are given by

\begin{equation}
\label{eq:Protogenovsolutions1}
rf_{\mu,s} 
=  
\frac{1}{gr}
\left[
1-\mu r\coth{(\mu r+s)} 
\right]\, ,
\hspace{1cm}
rh_{\mu,s} 
= 
\frac{1}{gr}
\left[
\frac{\mu r}{\sinh{(\mu r+s)}} 
-1
\right]
\, ,
\end{equation}

\noindent
and a 1-parameter family for which these functions, denoted by
$(f_{\lambda},h_{\lambda})$, are given by

\begin{equation}
\label{eq:Protogenovsolutions2}
rf_{\lambda} 
=  
\frac{1}{gr}
\left[
\frac{1}{1+\lambda^{2}r}
\right]\, ,
\hspace{1cm}
rh_{\lambda}
= 
-rf_{\lambda}\, .
\end{equation}

\noindent
The BPS 't~Hooft-Polyakov monopole \cite{Prasad:1975kr} is the only globally
regular solution and corresponds to $f_{\mu,s=0}$. The $f_{\mu,s=\infty}$
solution  is given by 

\begin{equation}
\label{eq:sinftysolution}
-rf_{\mu,\infty} =  \frac{\mu}{g} -\frac{1}{gr}, 
\hspace{1cm}
rh_{\mu,\infty} = -\frac{1}{gr}\, ,
\end{equation}

\noindent
and, for $\mu=0$, it is the Wu-Yang monopole \cite{Wu:1967vp}. The latter
solution is also recovered in the 1-parameter family for $f_{\lambda=0}$. 

The asymptotic behavior of $rf(r)$ (which is the combination that occurs in
the metrics we study) for the different solutions is

\begin{equation}
rf_{\mu,s} 
\sim  
-\frac{\mu}{g} +\frac{1}{gr} +\mathcal{O}(e^{-4\mu r})\, ,
\hspace{1cm} 
-rf_{\lambda}
\sim
\frac{1}{g\lambda^{2} r^{2}}+\mathcal{O}(r^{-3})\, ,  
\end{equation}

\noindent
and the behavior near the origin (where the black-hole horizons may be in the
metrics under study) are

\begin{equation}
rf_{\mu,0} 
\sim  
-\frac{\mu^{2}}{2g}r +\mathcal{O}(r^{3})\, ,
\hspace{.5cm} 
rf_{\mu,s} 
\sim  
\frac{1}{gr} -\frac{\mu}{g}\coth{s} +\mathcal{O}(r)\, ,
\hspace{.5cm} 
rf_{\lambda}
\sim
\frac{1}{gr} -\frac{\lambda^{2}}{g}r +\mathcal{O}(r^{3})\, .  
\end{equation}

If we define the magnetic monopole charge by 

\begin{equation}
p\equiv \frac{1}{4\pi}\int_{S^{2}_{\infty}}\mathrm{Tr}(\hat{\Phi} F)\, ,
\hspace{1cm}
\hat{\Phi} \equiv \frac{\Phi}{\sqrt{|\mathrm{Tr}(\Phi^{2})|}}\, ,
\end{equation}

\noindent
then, we always find $p=1/g$ except in the 1-parameter family for finite
$\lambda$, for which we find $p=0$. As we have argued in
Ref.~\cite{Bueno:2014mea}, the $\lambda\neq 0$ \textit{colored} monopoles can
be seen as a magnetic monopole placed at the origin whose charge is completely
screened at infinity.

%%%%%%%%%%%%%%%%%%%%%%%%%%%%%%%%%%%%%%%%%%%%%%%%%%%%%%%%%%%%%%%%%%%%%%
%%%%%%%%%%%%%%%%%%%%%%%%%%%%%%%%%%%%%%%%%%%%%%%%%%%%%%%%%%%%%%%%%%%%%%
%%%%%%%%%%%%%%%%%%%%%%%%%%%%%%%%%%%%%%%%%%%%%%%%%%%%%%%%%%%%%%%%%%%%%%
%%%%%%%%%%%%%%%%%%%%%%%%%%%%%%%%%%%%%%%%%%%%%%%%%%%%%%%%%%%%%%%%%%%%%%
%%%%%%%%%%%%%%%%%%%%%%%%%%%%%%%%%%%%%%%%%%%%%%%%%%%%%%%%%%%%%%%%%%%%%%
%%%%%%%%%%%%%%%%%%%%%%%%%%%%%%%%%%%%%%%%%%%%%%%%%%%%%%%%%%%%%%%%%%%%%%

\end{document}